\documentclass{emulateapj}

\begin{document}
\title{Observational Biases Masquerading as Cosmological Effects? A Cautionary Tale about Blue Tilts and Other Trends in Globular Cluster Systems \altaffilmark{1}}

\author{Arunav Kundu}

\affil{ Physics \& Astronomy Department, Michigan State University, East Lansing, MI 48824}

\altaffiltext{1} {Based on observations made with the NASA/ESA Hubble Space Telescope which is operated by
    the Association of Universities for Research in Astronomy, Inc., under NASA contract
    NAS 5-26555.}

\begin{abstract}
 
The high spatial resolution of the Hubble Space Telescope (HST) has led to tremendous progress in many areas of astronomy. The ability of the HST to peer into the bright inner regions of galaxies and distinguish between globular clusters (GCs) and background objects has been particularly beneficial to the study of clusters. But the very virtue of the HST that has been an asset 
 to such research can be its pitfall if the consequences of superior angular resolution are not considered in detail. Recent HST studies show a small, but consistent, color-magnitude correlation in the metal-poor halo GCs of nearby galaxies. This `blue-tilt' has been interpreted as a mass-metallicity relationship, implying self-enrichment in the higher mass GCs. We show that the `blue-tilt' is likely the consequence of a small, but measurable with HST, mass-size relationship in GCs. The combined effects of Poisson noise and surface brightness fluctuations can explain other apparent correlations of the  `blue-tilt' with environment. Some HST based studies have similarly suggested that the mean metallicity of the blue, metal-poor, halo clusters increases with host galaxy mass, indicating that GC metallicity is linked to the size of the host galaxy halo. We show that this correlation is also likely due to the effects of GC size in high resolution HST images. We also point out that the presumed fundamental plane of GCs itself varies with galactocentric distance due to GC size trends and ultra compact dwarfs may simply reflect the tail of the globular cluster distribution.

\end{abstract}


\section{Introduction}

The metal-poor globular clusters (GCs) in the halo of the Milky Way are among the oldest objects in the Galaxy. They have long provided crucial constraints on the preferred cosmological models of the day (e.g. Carroll  et al.~1992, Perlmutter  et al.~1997). Some recent studies starting with the currently favored cold dark matter paradigm suggest that such clusters, along with dwarf galaxies, may be the  stellar signatures of the rare peaks in the matter density function at z$\sim$12 (Moore  et al. 2006; Kravtsov \& Gnedin 2005).  GCs formed at such high redshift may even retain the spatial and kinematic properties of these early over-densities of baryonic matter (Parmentier \& Grebel 2005; Diemand, Madau, \& Moore 2005). There is an ongoing discussion on whether the formation process of these objects is affected by reionization, either by truncating the formation (Santos~2003; Bekki~2005), or in fact causing it (Cen~2001). Nevertheless it is generally agreed that halo globular clusters are among the charter stellar members of their host galaxies and provide a unique window into the histories of their hosts.

The globular cluster systems of nearby galaxies, especially cluster rich ellipticals, provide large statistically significant samples that allow increasingly sensitive tests of the galaxy assembly process. In recent years there has been an explosion of ever improving studies of the globular cluster systems of nearby galaxies (e.g. Kundu \& Whitmore 2001a, 2001b; Larsen et al. 2001; Rhode \& Zepf 2004). It has been established that most galaxies have bimodal GC color distributions (e.g. Kundu \& Whitmore~2001a; Peng  et al.~2006) consisting of a system of red GCs that roughly follows the bulge light, and a more dispersed halo population of blue GCs (e.g. Geisler, Lee, \& Kim 1996; Rhode \& Zepf 2004). The bimodal color distributions of GCs are thought to reflect peaks in the metallicity distributions, an interpretation that is confirmed by recent work (Zepf \& Ashman 1993; Kundu \& Zepf~2007; Strader  et al.~2007).

	Metal-rich GCs reveal many correlations with the properties of their host galaxies (e.g. Kundu \& Whitmore~2001a; Peng  et al.~2006; Strader et al. 2006) which are broadly consistent with the hierarchical merging scenario of galaxy, and associated globular cluster system, formation (e.g. Beasley et al. 2002). The effect of environment, if any,  on the metal-poor, halo clusters that have long been considered `primordial' objects (e.g. Peebles \& Dicke 1968) is less clear.  Ashman \& Bird (1993) first pointed out that metal-poor GCs appear to have remarkably similar characteristics in many galaxies of different types and sizes. While there is some evidence of cosmological bias in metal-poor clusters (Rhode, Zepf, \& Santos~2005) in the sense that large galaxies that began assembling earlier have more metal-poor GCs per unit mass, it appeared till fairly recently that the physical properties of these clusters were fairly universal (Forbes, Brodie, \& Grillmair 1997; Kundu \& Whitmore 2001a). 

Among the more intriguing suggestions made by recent Hubble Space Telescope - Advanced Camera for Surveys (ACS) studies of globular cluster systems is the existence of a `blue-tilt', or a mass-metallicity relation in GCs. Several groups have independently reported the tendency of blue, metal-poor GCs to drift to redder colors with increasing GC luminosities (Strader  et al.~2006; Harris  et al.~2006; Mieske  et al.~2006; Spitler  et al.~2006). While some studies suggest a smaller, but measurable, trend in the same direction for the metal-rich subsystem (Spitler et al. 2006), others do not see such an effect (Strader et al. 2006).  The various groups generally agree that the `blue-tilt' in the metal-poor GCs is likely due to self-enrichment of large GCs, although they differ on whether this is because these clusters had dark matter halos that have since been stripped (Strader et al. 2006), or if they formed in large gas clouds (Harris et al. 2006). Other studies have attempted to study the theoretical basis for this effect (e.g. Bekki, Yahagi, \& Forbes 2007; Rothberg et al. 2008).

A number of correlations have been discovered between the `blue-tilt' in GCs and their host galaxies. The mass-metallicity trend appears to be stronger in more massive galaxies, and especially so in the inner regions (Mieske  et al.~2006). These papers also agree of that the `blue-tilt' is more apparent when the color-magnitude distribution of GCs is plotted using the redder  of the two filters on the magnitude axis, which has been assumed to be the consequence of the superior mass sensitivity of redder filters. In fact these trends are so continuous that there even appears to be a `red-tilt' in the color-magnitude distribution of the {\it metal-rich} GCs in the least massive galaxies in the Mieske et al. (2006) sample when the color-magnitude diagram is plotted with the blue filter on the magnitude axis.

	Another interesting discovery with potential cosmological significance is the trend of increasing mean metallicity of GCs in more massive galaxies. While all globular cluster studies agree about the existence of a strong link between the galaxy mass and GC metallicity for metal-rich GCs, the evidence for a similar, albeit weaker, correlation in metal-poor GCs was less clear in previous data sets (e.g. Kundu \& Whitmore~2001a; Larsen  et al.~2001). The recent studies of Peng  et al.~(2006) and Strader  et al.~(2006) propose a mean metallicity trend for the metal-poor GCs. Strader, Brodie, \& Forbes~(2004) suggest that this points to an in situ picture of galaxy formation with the biased self-enrichment of larger halos accounting for the correlation of the mean colors of blue, metal-poor GCs with galaxy mass. 
  
	We note that although both of these trends are very small, $\sim$0.1 mag in color from the brightest to the faintest objects, the various published studies on the variations in the colors of the metal-poor GCs argue that they are statistically significant. In this paper we investigate several subtle observational effects overlooked in these studies in order to answer the basic question of just how universal the properties of the `primordial', metal-poor halo globular clusters are. In particular we investigate the effect of cluster sizes on the photometry and comment on the effects on related issues such as the globular cluster luminosity function and the attempt to identify transition objects between GCs and dwarf galaxies. 

\section {Results \& Discussion}

\subsection {To Be or Not to Be? The `Blue-Tilt' in M87 Revisited }

M87 the giant elliptical at the center of the Virgo cluster is the galaxy with the richest known globular cluster system, and has long been a touchstone of globular cluster research. Not surprisingly, M87 is one of the galaxies in which the `blue-tilt' phenomenon was discovered. Using the same set of ACS-WFC images of M87 obtained in the F475W (g) and F850LP (z) filters as part of HST program GO-9401 (PI: Cote), Strader et al. (2006) and Mieske et al. (2006) showed that M87 provides the strongest evidence for this phenomenon. In this section we reanalyze these images and reassess the `blue-tilt'.

	The standard ACS pipeline reduced multidrizzled F475W and F850LP band HST GO-9401 program images of M87, observed on 2003, Jan 19$^{th}$,  were obtained from the HST archive. From hereon we refer to these images, and the associated photometry, as the g and z bands respectively as a shorthand for the bands these filters nominally represent. Globular cluster candidates were identified in the ACS images using the constant signal-to-noise ratio technique described in Kundu et al. (1999). We compared our lists to that of the Strader et al. (2006) study (kindly provided to us by Jay Strader) and found excellent agreement between the two. In the following analysis we use our own list, but note that all of the trends and conclusions discussed below are unchanged, irrespective of the particular list that is considered.

	Since the method of measuring the light from a globular cluster is at the heart of much of the subsequent discussion we use the following convention throughout this paper: We define the sky value around each object as the median of the pixels that are within the annulus between 5 and 8 pixels radius of the source. When we state that aperture photometry is performed within  a fiducial aperture, e.g. 1 pixel, we mean that the flux within a 1 pixel radius aperture is measured, the background counts based on the sky definition above are subtracted, {\it an aperture correction is applied} to calculate the total counts in the source, and the counts are converted to magnitudes using the appropriate photometric zero point. On the other hand, when we mention that the light within an aperture, say 2 pixels, is measured we mean that the flux within a 2 pixel radius aperture is measured, the background counts based on the sky definition above are subtracted, and the counts are converted to magnitudes using the appropriate photometric zero point, {\it with no aperture correction applied.} Throughout this paper a number in the subscript associated with a photometric magnitude refers to the radius of the aperture used for photometry. For example, z$_{0.5}$ refers to aperture photometry performed with an aperture of radius 0.5 pixels, a background or sky value defined by the median of the pixels in the annulus between 5 and 8 pixels radius, and appropriate aperture correction and zero points. We also note that any reference to the size of an aperture always refers to the radius in pixels, and not the diameter.

	We performed aperture photometry for the sources identified by our   constant signal-to-noise ratio technique using a 3 pixel radius aperture in both the g and the z filter. The ABMAG zero points and aperture corrections published by Sirianni et al. (2005) were used to obtain the colors and magnitudes of the cluster candidates. We note that all colors and magnitudes in this paper are in the ABMAG system. We also point out that the aperture corrections in Sirianni et al. (2005) are for point sources, and an additional small aperture correction is needed to get accurate absolute photometry of partially resolved GCs in HST images, and to account for the fraction of the source flux that falls within the sky annulus (e.g. Kundu et al. 1999). However, as will become clearer later in the paper, just what the choice of such a correction should be has a crucial effect on the very trends being studied in this paper. Therefore we use the point source corrections of Sirianni et al. (2005) for the baseline analysis in order to probe the {\it relative} variation of GC colors and magnitudes, and the effect on the {\it structure} of the color-magnitude diagram (CMD).

	The final M87 images from the HST GO-9401 program were created from two independent images in each filter (as is the case for all the other galaxies in this program) in order to remove the effects of cosmic rays, chip defects etc. 
Clearly some fraction of the pixels in many of the cluster candidates are flagged in one of the two images, and the signal in these pixels is contributed by only one image. In cases where a masked pixel corresponds to the central pixel  of a point-like source , which contributes a substantial fraction of the total light, the actual noise may be significantly larger than that calculated by Poisson statistics assuming uniform weights for all pixels. Moreover, the ACS camera is an off-axis instrument on the HST with significant geometric distortion across the WFC chips. Thus the final geometrically corrected image is created by combining the two slightly offset dithered images using the drizzle algorithm (Fruchter \& Hook 2002). In instances where the central pixel is masked in one of the images the center of a point-like source can shift, leading to photometric errors. Furthermore, in the few cases where the central pixel of a point-like source in both input images is masked out the drizzle algorithm may construct a final image where the light in the central pixel is contributed entirely by surrounding pixels with lower counts, leading to significant photometric errors. 

Fortunately the drizzle algorithm creates a weight image that keeps track of the number of pixels that contribute to the signal in each final drizzled pixel. Input pixels that are masked due to cosmic rays, hot pixels, decoding errors etc. are given zero weights in such images. In order to account for these effects we reject `low weight' cluster candidates for which the weight of the central pixel is more than 0.1 counts below that of the average value of the surrounding pixels. A cutoff value of 0 was not chosen because there is spatial variation in the weight images due to the geometric distortion of the ACS. Of course the effects described above contribute additional noise for pixels other than the central one in a point-like globular cluster, but since these effects are significantly larger for the innermost pixel we only account for this.

\begin{figure}
\includegraphics[angle=0, scale=0.6]{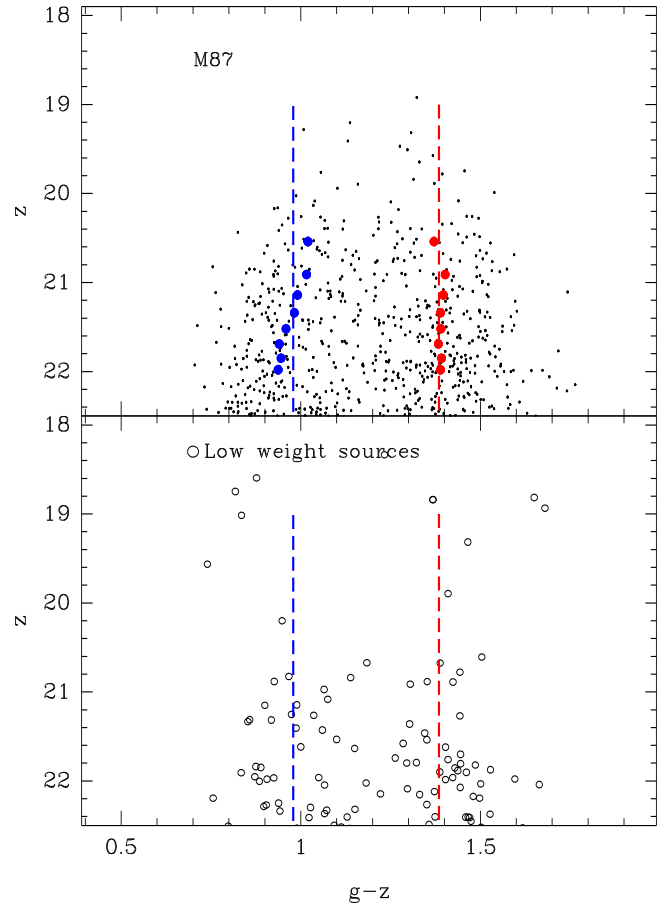}
\caption{ Top: g-z vs. z color-magnitude diagram of globular clusters in HST-ACS images of M87. The dashed lines mark the peaks of the blue and red globular cluster sub-populations. The large filled circles indicate the locations of the peaks of the red and blue sub-populations of a rolling sample of 150 clusters (see text for details). As in previous studies there is a color-magnitude trend for the blue clusters, with brighter blue GCs appearing redder, while there is no such trend for the red GCs. Bottom: Cluster candidates that happen to lie at locations where the central pixel is masked out in one or more of the input images. These objects are not included in the study because the larger color and magnitude uncertainty can skew the results. }
\end{figure}

The top panel of Fig 1 plots the g-z vs. z color-magnitude distributions of the cluster candidates in M87 which have colors between 0.6$<$g-z$<$1.9 and formal Poisson uncertainties of less than 0.1 mag in each filter.  The bottom panel shows the color-magnitude diagram of the `low weight' GCs that are rejected using the method described above. The bright `low weight' GCs clearly show a larger color spread than candidates in the clean GC sample, likely because of the issues discussed above.  It is possible that some of the low weight sources that we reject in fact have reliable colors and magnitudes. Since our rejection method is unbiased with respect to the location and/or brightness of a GC, and does not affect the measurement of a `blue-tilt', we choose to analyze the clean sample.

The well known bimodality in the GC distribution of M87 is apparent in Fig 1. We performed KMM mixture modeling tests (e.g. Ashman, Bird, \& Zepf 1994) in order to quantify the bimodality and measure the peaks (both here and throughout the rest of the paper). The homoscedastic test (which assumes that the sub-populations have identical variances) confirms that the distribution is bimodal. Although the results in this paper are unaffected if the heteroscadastic test (in which the variances are allowed to vary) is applied, we opt to consistently use the homoscedastic test throughout this paper because the added uncertainty introduced by an extra free parameter simply increases the noise for some of the smaller samples in this study. The dotted lines indicate the peaks of red and blue sub-populations for the 500 brightest sources. The large dots indicate the locations of the peaks of the red and blue populations of a rolling sample of 150 successive GC candidates, sampled every 50th cluster from a luminosity sorted list of the 500 brightest GCs. As in previous studies an apparent color-magnitude trend is observed in the blue GCs, while there is no obvious trend for the red GCs. 

We note that we restrict this analysis to the 500 brightest clusters because the various independent studies agree on the existence of this feature for the brightest GCs but disagree about the continuation of the effect to fainter magnitudes (Harris et al. 2006; Strader et al. 2006; Spitler et al. 2006). Moreover, even though Strader et al. (2006) suggest that the `blue-tilt' in M87 extends to lower luminosity clusters they reject the formal fit at one of the fainter magnitudes which is inconsistent with the trend because of an apparent `H' structure in the color-magnitude diagram. Since the evidence for (or against) the `blue-tilt' hinges on the behavior of the most luminous highest S/N sources which are presumed to have the most secure photometry, and the contamination from background and foreground sources is negligible for the bright GCs (e.g. Kundu et al. 1999; Peng et al. 2006), we concentrate on these high mass sources.

	We perform aperture photometry in both the g and z images for aperture radii of 0.5, 1, and 4 pixels. A few clarifications on the choice of aperture corrections are in order at this juncture. As mentioned above we adopted the point source aperture correction from Sirianni et al. (2005) for the 3 pixel aperture photometry.  But as the GCs are slightly resolved this ignores a small additional term (plus any small offset due to the fraction of the point spread function that falls within our sky region). However, if the GCs are unresolved, and/or have no discernible mass-radius relation as is often claimed (e.g. Jordan et al. 2005), this is simply a constant offset that applies to all GCs and would not affect the structure/shape of the CMD. Adopting the Sirianni et al. (2005) 0.5, 1, and 4 pixel aperture corrections for the respective photometric apertures would similarly underestimate the aperture correction. Moreover it would be underestimated by different amounts in apertures of different sizes. This would introduce zero point offsets between the photometry with different size apertures complicating our goal of studying the effect of photometric aperture size on the structure of the color-magnitude diagram. Therefore, we have chosen to adopt the 3 pixel aperture photometry as the reference set. We then measure the difference between the light within 0.5 pixels and the 3 pixel aperture photometry and adopt this as the sum of the zero point and aperture correction for the 0.5 pixel photometry. The 1 and 4 pixel photometry were calibrated similarly. We tested for variations in these corrections by comparing in turn the entire set of GCs, a handful of the brightest sources, and sources with small photometric uncertainties. There was no significant difference in the absolute magnitude calibration of the 0.5, 1 and 4 pixel aperture photometry using the various sets of comparison sources. 

\begin{figure}
\includegraphics[angle=0, scale=0.9]{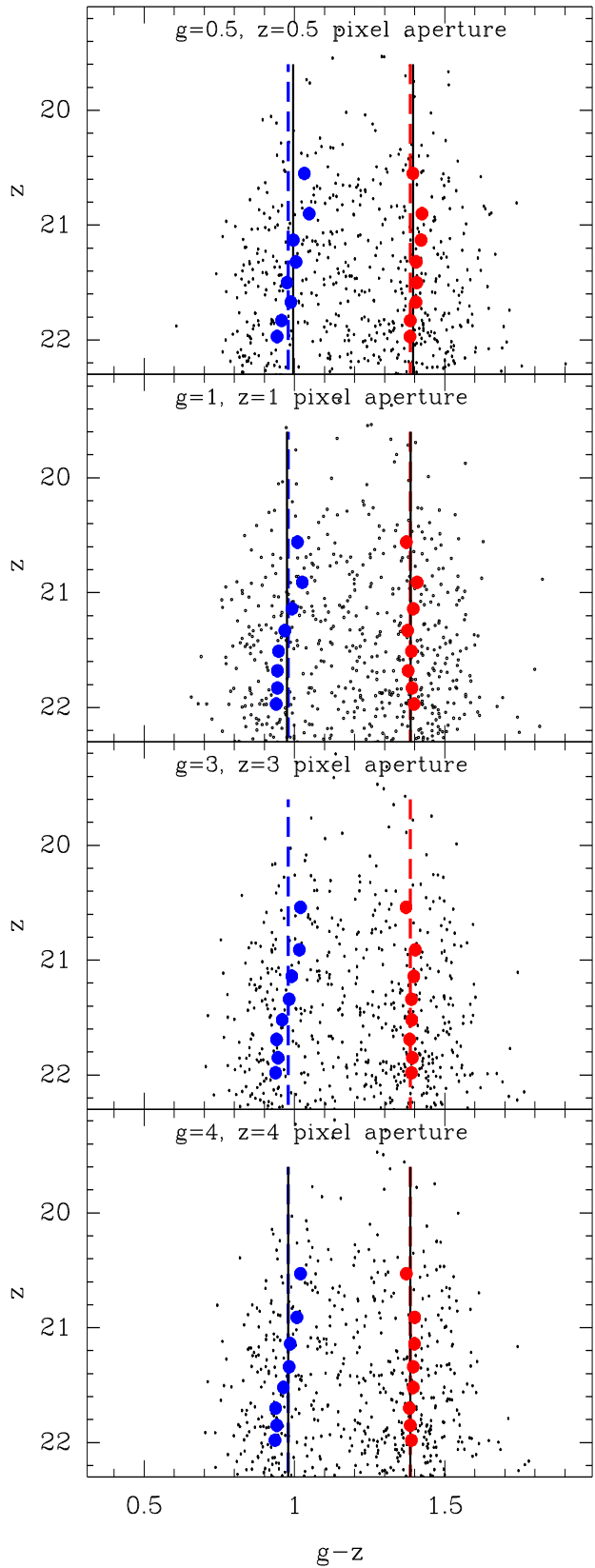}
\caption{ Color-magnitude diagrams of GCs in M87 derived using apertures of 0.5, 1, 3, \& 4 radius in both the g and z filters. Symbols as in Fig 1. The dashed lines in each panel indicate the blue and red peaks measured for the 3 pixel aperture photometry. In each case there is a `blue-tilt' in the blue GCs and no such trend in the red clusters. The location of the peaks and the magnitude of the `blue-tilts' appear similar in all four plots. }
\end{figure}

\begin{figure*}
\includegraphics[angle=-90, scale=0.8]{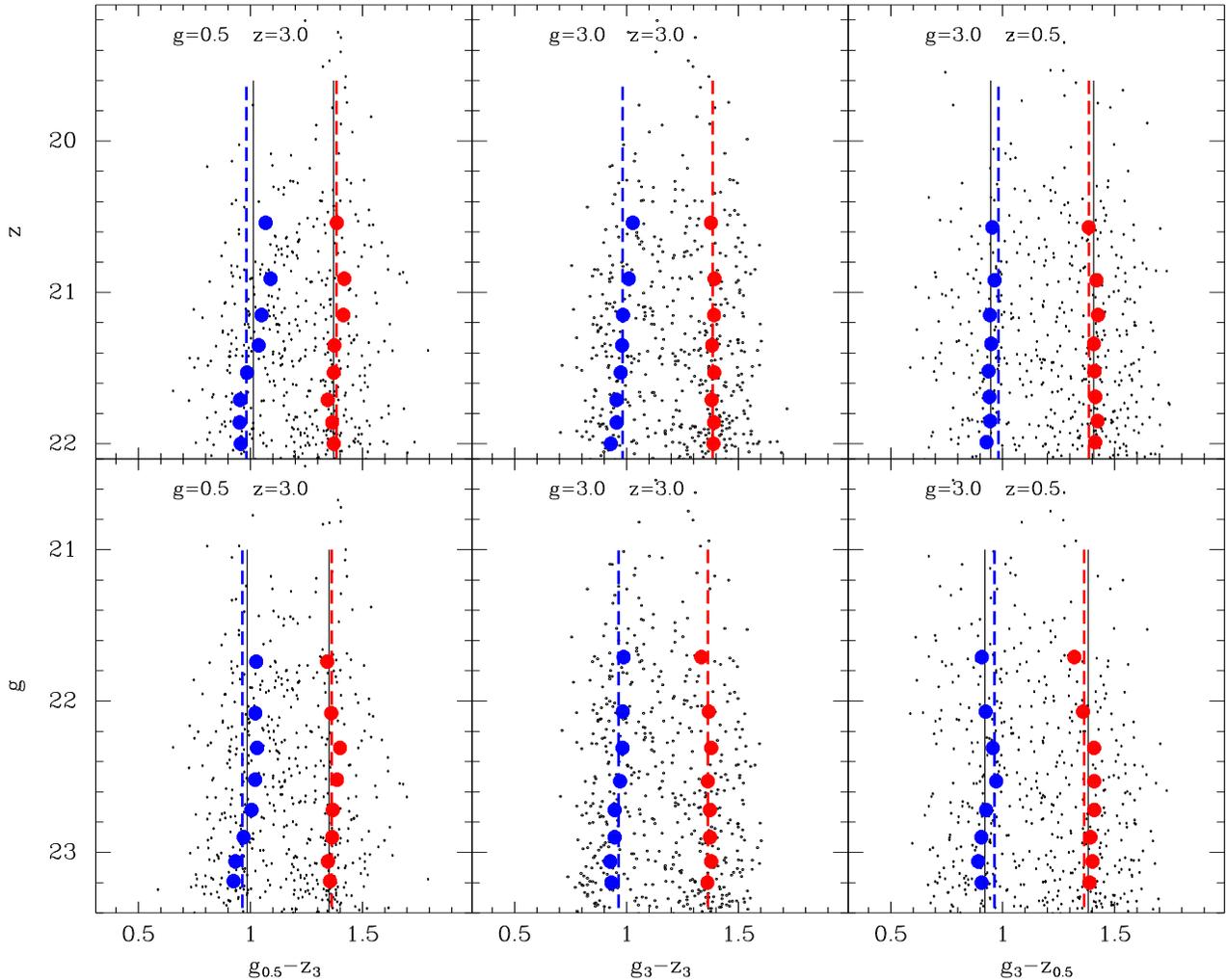}
\caption{ Top row: The g-z vs z CMDs of globular clusters in M87 with different combinations of aperture radii. The dashed lines indicate the peaks of the blue and red GCs measured in the g$_3$-z$_3$ vs. z$_3$ data set, while the solid lines indicate the peaks in the data plotted in the particular panel. The `blue-tilt' is strongest in the g$_{0.5}$-z$_3$ vs. z$_3$ CMD and weakest in the g$_3$-z$_{0.5}$ vs. z$_{0.5}$ CMD. Bottom row: The corresponding g-z vs. g CMDs for different aperture combinations. As in previous studies the `blue-tilt' appears weaker in the g-z vs g CMDs as compared to the  g-z vs. z plots.  While the top left image shows the strongest `blue-tilt' the bottom right figure suggests a `{\it red}-tilt'. }
\end{figure*}

Fig 2 shows the g$_{0.5}$-z$_{0.5}$ vs. z$_{0.5}$, g$_{1}$-z$_{1}$ vs. z$_{1}$, g$_{3}$-z$_{3}$ vs. z$_{3}$, and g$_{4}$-z$_{4}$ vs. z$_{4}$ color-magnitude distributions of the M87 GCs using different photometric apertures. The KMM determined peaks of the blue and red populations of the 500 brightest clusters in each set are also marked. As in Fig 1 the blue and red peaks for a rolling subset of 150 GCs from KMM mixture modeling are shown. Overall the structure of the color-magnitude distributions shows little variation with the choice of aperture. There is a negligibly small variation in the overall peaks in the four plots, although there is a slight hint that the `blue-tilt' may be stronger for the smallest aperture pair. In fact the g$_{0.5}$-z$_{0.5}$ vs. z$_{0.5}$ photometry shows a slight `blue-tilt' even in the metal-rich GCs.

	Next we combine the photometry from different combinations of apertures, with strikingly different results. The top panels of Fig 3 plot the g$_{0.5}$-z$_{3}$ vs. z$_{3}$, g$_{3}$-z$_{3}$ vs. z$_{3}$, and g$_{3}$-z$_{0.5}$ vs. z$_{0.5}$  CMDs, while the bottom panels show the corresponding g$_{0.5}$-z$_{3}$ vs. g$_{0.5}$, g$_{3}$-z$_{3}$ vs. g$_{3}$, and g$_{3}$-z$_{0.5}$ vs. g$_{3}$ CMDs.  Consistent with previous observations, the middle panels  show that the `blue-tilt' is stronger in the g-z vs. z plane as compared to the g-z vs. g plane. However the top panels show that the `blue-tilt' in the metal-poor GCs increases when a small aperture is chosen in the g band and a large one in the z band. There is even the hint of a `blue-tilt' in the metal-rich GCs in the top left figure (g$_{0.5}$-z$_{3}$ vs. z$_{3}$ CMD). On the other hand the `blue-tilt' is not obvious in either sub-population of GCs in the top right panel where a large g, and small z aperture has been chosen (g$_{3}$-z$_{0.5}$ vs. z$_{0.5}$). The lower set of panels of g-z vs. g plots shows a similar trend, except that for each combination of apertures it reveals less of a `blue-tilt' than the corresponding g-z vs. z CMDs. In fact the trends are so continuous that the g$_{3}$-z$_{0.5}$ vs. g$_{3}$ CMD indicates a `red-tilt' for the bright GCs, especially for the metal-rich clusters. The difference in the structure of the CMDs of the brightest GCs is dramatically different between the top left and bottom right figures. While the bright GCs tend to lie on a curve from the bottom left to the top right in the former plot (g$_{0.5}$-z$_{3}$ vs. z$_{3}$ CMD), they are roughly lined up from the bottom right to the top-left in the latter (g$_{3}$-z$_{0.5}$ vs. g$_{3}$ CMD). In other words even for the brightest candidates, which are presumed to have the most reliable photometry, the `blue-tilt' can be turned into a `red-tilt' 
based on the choice of photometric measurement parameters. Fig 3 also indicates that the peak colors of the overall metal-rich and metal-poor populations drift based on the choice of apertures. We investigate the underlying causes below.

\subsection {One Size Doesn't Fit All! The Consequences of a Mass-Size Relationship}

Globular clusters in the Milky Way have a mean half-light radius of $\approx$3 parsecs. Various studies have shown that such objects are partially resolved in HST images at Virgo Distances (e.g. Kundu \& Whitmore 2001a, 2001b; Larsen et al. 2001; Jordan et al. 2005). A simple way to check if a point-like source is actually resolved, that has often been employed, is to measure the difference in light within two apertures of different sizes. A combination that works well at the resolution of the ACS-WFC, which is similar to the PC chip of the WFPC2, is  calculating the difference in light between a 0.5 and 3 pixel aperture, or $\Delta{(0.5-3)}$ (e.g. Kundu et al. 1999). In this scheme compact objects that have a larger fraction of the total light within 0.5 pixels have smaller $\Delta{(0.5-3)}$ values while more diffuse objects have larger $\Delta{(0.5-3)}$ values.

\begin{figure}
\includegraphics[angle=-90, scale=0.4]{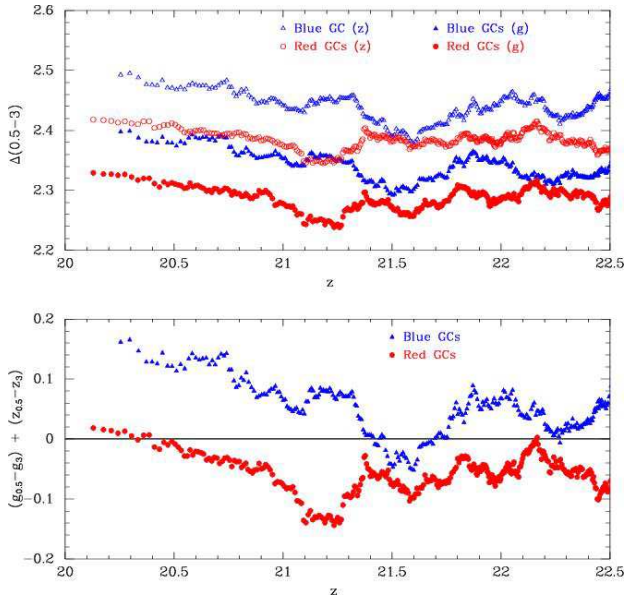}
\caption{Top: The difference in the light measured within 0.5 and 3 pixel apertures, $\Delta$(0.5-3), as a function of z band magnitude. The running average of the $\Delta$(0.5-3) of 35 clusters is plotted for both the red and blue GCs in the g and z filters. $\Delta$(0.5-3) traces the effective size of the GCs. There is a small, but measurable tendency for the brighter GCs to be larger. The apparent structure in the size distribution of fainter GCs reflects the limitations set by the poor sampling of these particular ACS images (see text for details). Bottom: The (g$_{0.5}$-g$_3$) + (z$_{0.5}$-z$_3$) magnitudes of the blue and red GCs as a function of z. Note that  (g$_{0.5}$-g$_3$) + (z$_{0.5}$-z$_3$) is identical to the sum of the g and z band $\Delta$(0.5-3) values {\it with the appropriate aperture corrections applied}, and hence the sum of the g and z band size estimates.  The plot traces the difference between the peaks of each, the red, and the blue GCs in the left and right panels of Fig 3. The difference in the tilts can thus be explained mainly by size differences. Note that the average (g$_{0.5}$-g$_3$) + (z$_{0.5}$-z$_3$) value of the blue GCs is a positive number while  that of the red ones is a negative number. This is driven by the larger sizes of the blue GCs as compared to the red clusters and reflects drift in the overall peaks in Fig 3 (see text). }
\end{figure}

	In the top panel of Fig 4 we plot the running average of the $\Delta{(0.5-3)}$ values for the metal-rich and metal-poor subsystems of GCs in M87 in the g and z filters as a function of z-band magnitude. We note that both in this figure, and throughout the rest of the paper we have used the threshold color returned by the KMM test to divide the sample into red and blue globular cluster subsystems. Each point in Figure 4 represents the average of 35 GCs. As we first pointed out in Kundu \& Whitmore (1998), and has since been well established in many subsequent studies (e.g. Larsen et al. 2001; Jordan et al. 2005; Spitler et al. 2006), the metal-poor, blue globular clusters are larger in size than the metal-rich, red ones. The z band curves have larger $\Delta{(0.5-3)}$ magnitudes because the point spread function in the z image  has a larger FWHM. However, the resolving power of diffraction limited telescopes such as the HST is higher at shorter wavelengths. 

	An interesting feature of Fig 4 is that for the brightest few magnitudes there is a clear correlation between the cluster luminosity and the $\Delta{(0.5-3)}$ size in both filters for both the metal-rich and metal-poor GCs. In other words there is a mild (but measurable) mass-size relationship for M87 GCs. The globular clusters in the Milky Way have long been known to have half light radii of $\approx$3 pc with a puzzling lack of a significant mass-radius relationship (Djorgovski \& Meylan 1994; van den Bergh 1996). While extragalactic cluster systems show a similar mass-radius structure (Kundu et al. 1999), this does not imply that these quantities are completely uncorrelated, as for example assumed by Jordan et al. (2005). Studies of the old globular clusters in M87 and NGC 4594 (Waters et al. 2006; Spitler et al. 2006) suggest a weak $\sim$L$^{0.04-0.09}$ correlation between mass and radius. The clusters in the Milky Way and other local group galaxies are consistent with this mild trend, especially at the high mass end (Barmby et al. 2007). 

We note that this trend continues to higher GC masses in M87 but because of the combination of small number statistics in the tail of the globular cluster luminosity function (GCLF) and the fact that we only plot the running average of 35 sources it is not seen in Fig 4. The mass-size relationship induces subtle, but measurable biases in the color-magnitude diagrams of clusters. When a fixed aperture correction is used for clusters of all sizes the total luminosity of larger objects is underestimated. This bias is larger in the bluer filter where the objects are better resolved due to the more compact PSFs at shorter wavelengths in diffraction limited HST images. Thus larger objects appear redder. Since the metal-poor GCs are on average larger and better resolved than metal-rich clusters this apparent `blue-tilt' effect is stronger for the blue clusters. When the photometry for a large z aperture is combined with that of a small g aperture the effect of such a size resolution on the CMDs is enhanced, such as in the left plots of Fig 3. On the other hand when a larger g aperture is combined with a smaller z aperture there is relatively less of a size resolution in the bluer filter and the fraction of light sampled in the two filters are better correlated for objects of all sizes. Hence the weakening of the blue-tilt in the right panels of Fig 3. Another subtle issue is that because of the correlation of GC size with color and magnitude the objects at the top left of the CMD are the largest ones; Hence the luminosity of these objects is underestimated by the largest amount when uniform aperture corrections are applied. This can further enhance the `blue-tilt' in the g-z vs. z CMDs (and in fact weaken possible red tilts in the right panels of Fig 3).

We now digress slightly to comment on the mass-size relation at fainter magnitudes and the believability of the apparent structure observed for the less luminous GCs in Fig 4. Since the `blue-tilt' is primarily seen in high mass GCs this particular discussion does not affect the arguments pertaining to the `blue-tilt'. First we note the mass-size relation seen in the luminous GCs likely reflects an initial trend because dynamical evolution due to evaporation, particularly at larger galactocentric distances where tidal shocks are less important, is expected to induce the opposite trend, especially for lower mass GCs (Fall \& Rees 1977). The structure in the apparent size distribution for the low luminosity clusters,  particularly the smaller red GCs fainter than z$\sim$21.2 is likely not reliable to test for the effects of evaporation. This is because the GO-9401 images are constructed from just two, dithered ACS images that have high geometric distortion, undersample the point spread even in drizzled images, and do not fully account for spatial variations in the PSF (Anderson \& King 2006). In such a situation the exact location of the peak of a GC within the central pixel has significant influence on the distribution of light and hence on the derived size. This is obviously more important for smaller GCs that are closer to the resolution limit of the HST. Moreover in both the Milky Way (van den Bergh 1996) and other galaxies (Kundu et al. 1999) the range of GC size distributions increases at lower mass. Thus for lower luminosity objects the size of a larger fraction of GCs, especially the smaller red clusters, is unresolved in these ACS images and biases the average. And finally the lower signal-to-noise in less luminous GCs of course makes size estimates more unreliable. The analysis of a deeper M87 data set in the next section, which shows far less structure in the size-magnitude relation of fainter GCs suggests that these sources of noise in the poorly sampled GO-9401 g and z images induce the apparent size upturn at lower luminosities in Fig 4.  

We note that since the mass-size relation is only apparent in the bright clusters that comprise the tail of the GCLF, any attempt to fit a linear relationship that includes fainter, and more numerous GCs which have less size information will likely yield no formal correlation between the mass and the size of the GCs. This is likely the reason why Jordan et al. (2005) find an overall mass-size slope consistent with zero for all sources brighter than the GCLF turnover. On the other hand the GCs in the nearby galaxy NGC 4594 are better resolved in HST-ACS images and Spitler et al. (2006) find an obvious mass-size trend in the most luminous blue GCs, and a shallower trend in the red GCs. Consistent with our explanation above Spitler et al. (2006) find a significant `blue-tilt' in the blue GCs, and a smaller, but consistent `blue-tilt' even in the red GCs.

In the bottom panel of Fig 4 we plot the rolling average of the (g$_{0.5}$-g$_3$) + (z$_{0.5}$-z$_3$) magnitudes of the blue and red GCs as a function of z. Note that  (g$_{0.5}$-g$_3$) + (z$_{0.5}$-z$_3$) is identical to the sum of the g and z band $\Delta$(0.5-3) values {\it with the appropriate aperture corrections applied}, and hence the sum of the g and z band size estimates.  This plot effectively shows the difference between the color-magnitude trends in  the metal-rich and metal-poor populations in the g$_{0.5}$-z$_{3.0}$ vs. z$_{3.0}$ CMD as compared to the g$_{3.0}$-z$_{0.5}$ vs. z$_{0.5}$ CMD in Fig 3.  The significant difference in the slope of the `blue-tilt' for the luminous sources, especially the metal-poor ones, is apparent. The lower panel of Fig 4 also shows that the average offset in the metal-poor GCs is positive, while that of the metal-rich ones is negative, albeit of a smaller amplitude.  This is driven by the larger sizes of the blue GCs as compared to the red clusters and reflects drift in the overall peaks of the red and blue GCs in various panels of Fig 3.
 As the smaller metal-rich clusters outnumber the larger metal-poor ones in HST studies of  M87 (e.g. Kundu et al. 1999; Peng et al. 2006) the mean shifts for the metal-rich GCs in the upper panel of Fig 4, and consequently the shifts in the mean colors of the metal-rich GCs in the various panels of Fig 3, are smaller for these clusters. It is clear however that the peak colors of metal-rich and metal-poor peaks for individual galaxies can vary based on the mean half-light radii of the various sub-populations. We discuss the implications of this in $\S$2.5.

\subsection {What's in a Name? Uncertainty by any Other Name is Still an Uncertainty}

We have shown that the `blue-tilt' is likely caused by a small, but measurable, mass-size correlation in globular clusters. Now we investigate the reasons for the differences in the magnitude, and possibly the direction, of this phenomenon with the choice of filters and galaxies.

	The observed structure in globular cluster CMDs can be affected by uncertainties in the photometry of individual GCs. However, the formal Poisson uncertainties for the bright sources in M87 are negligibly small. Fig 3 of Strader et al. (2006) for example shows that the brightest sources have photometric uncertainties of less than 0.01 mags. As the flat fielding errors in the ACS chips alone are of the order of 1\%, these estimates (of Strader et al. 2006 and the Poisson estimates in all other studies, including this one) are lower limits. It is worth recalling that the superior angular resolution of the cameras on the HST has enabled us to resolve out more of the diffuse galaxy light and push the study of GCs closer to the center of the galaxy. The bright galaxy background in such studies contributes significant surface brightness fluctuation noise. Such astronomical noise is convolved by the PSF and is not estimated by the standard CCD noise formula. The background noise in ACS images is further complicated by the fact that adjacent pixels, especially in poorly sampled drizzled images such as the ones analyzed above, are correlated.

\begin{figure*}
\includegraphics[angle=-90, scale=0.7]{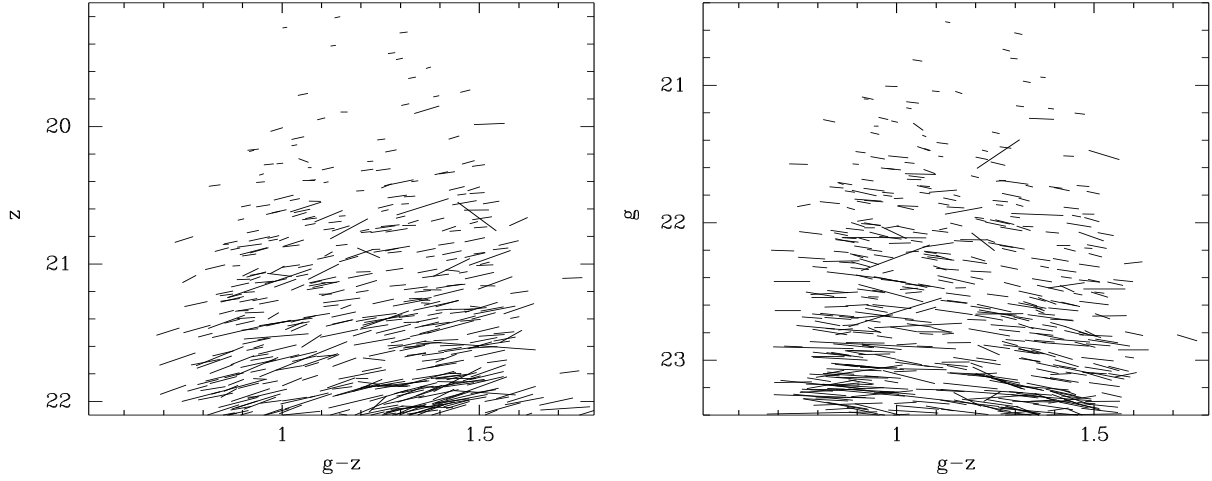}
\caption{ The uncertainty in the g-z vs. z and g-z vs. g photometry for each GC candidate due to the photometric uncertainty in each filter (See text for details). As explained in the text the direction of the correlated color and magnitude uncertainties tends to magnify the `blue-tilt' when plotting the luminosity using the z filter and suppress it when the g filter is considered.}
\end{figure*}

\begin{figure*}
\includegraphics[angle=-90, scale=0.9]{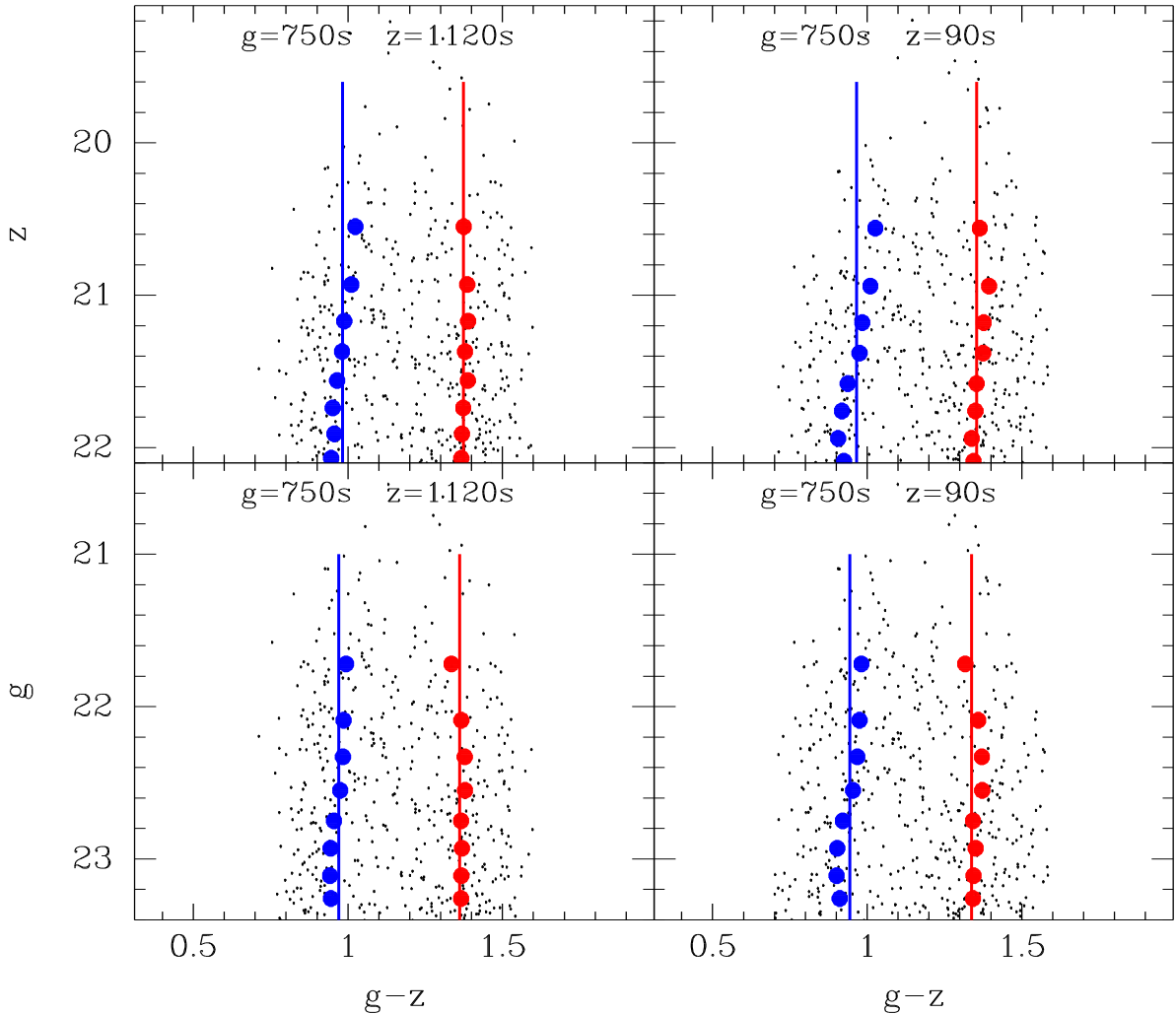}
\caption{ CMDs of M87 GCs using both the deep 1120s and shallow 90s z band images. The `blue-tilt' is clearly larger with the shallow z image (for both the blue and red GCs) revealing the effects of observational noise on the apparent trend. }
\end{figure*}

In order to empirically estimate the background noise we calculated the flux within a 3 pixel aperture at 8 locations, 25 pixels from the center of each GC candidate. We adopted the same 5-8 pixel sky annulus as for the GC photometry. To each of this set of 8 background flux measurements for a GC candidate we added the Poisson sampled background subtracted counts for the corresponding object i.e. the background subtracted counts are assumed to follow a Poisson distribution and we randomly drew from such a distribution to preserve the Poisson noise component.  We then converted the total counts corresponding to each background location to magnitudes using the zero points and aperture corrections described in $\S$2.1. The range of magnitudes traced using this set of 8 measurements for each GC gives a more realistic estimate of the photometric uncertainties. In the following discussion we only consider sources for which the maximum separation of the extreme values of the g or z magnitudes (as appropriate) in the 8 sky measurements is less than 0.3 mags. This minimizes the effect of artificially large noise when a simulated sky location happens on fall on another point source, bad pixel, the edge of the drizzled image, etc. 

In Fig 5 we select the two sky locations for the measurements with the largest difference in g-z color for each GC and connect the points on g-z vs. z and g-z vs. g plots. While these lines are formally not the exact uncertainties because they reflect the maximum separation from just 8 trials, they do give a general sense of the magnitude of the uncertainty. It is clear that the actual uncertainties in the measurements are larger than the formal Poisson errors. Since the color of a GC has a component of magnitude the uncertainties in the two axes of a CMD are obviously correlated. Fig 5 shows that the uncertainty causes GCs to drift from the bottom left to top right in the g-z vs. z plot (or vice-versa), {\it hence reinforcing the `blue-tilt'}, and from the bottom right to top left in the g-z vs. g plot, {\it thus weakening the `blue-tilt'}. Fig 5 also reveals that the slope of the error changes mildly with color such that the `blue-tilt' is amplified for metal-poor GCs in g-z vs. z CMDs while a corresponding `red-tilt' is induced in the metal-rich GCs in g-z vs. g plots. This is because for GCs at the same z magnitude in the g-z vs. z CMD the redder, more metal-rich GCs are fainter in g. The correspondingly larger g (and g-z) uncertainty changes the direction of the uncertainty ellipse (and vice versa for the blue CMDs). We note that such variations in the uncertainty ellipses that may affect the structure of the underlying color-magnitude distributions are a generic property of any CMD and should be carefully accounted for whenever the noise in the individual measurements is significant e.g. when attempting to decipher the age and metallicity structure of distant galaxies by probing the color and magnitude distribution of individual stars.

 The effect of uncertainties on the `blue-tilt' can be directly studied by comparing data sets with different exposure times. The GO-9401 observations of M87 also included a 90s exposure in the z-band. We measured the magnitudes of the GCs in this image using the exact same photometric parameters as used in Fig 1. Fig 6 presents the CMDs of the clusters using different combinations of filters and exposure times, along with the peaks of the red and blue GCs. Clearly there is more of a `blue-tilt' in the metal poor GCs in the g-z vs z CMD with the shallower z data. The g-z vs g CMDs show a similar, but weaker, trend due to the moderating effects of the uncertainties in the GC photometry on the `blue-tilt' described above.

The same experiment can be conducted by analyzing a much deeper data set. To this end we analyzed the deep F606W (V) and F814W (I) images of M87 obtained with the HST-ACS camera as part of HST GO program 10543 (PI: Baltz). We downloaded sets of dithered images and co-added the individual exposures to create a single deep 17,000s V image and a 50,400s I image. We note that this represents the portion of the data uncorrupted by telescope or instrumental errors that was available publicly at the time of this particular analysis. Including additional data that has since become public does not affect any of our conclusions.

\begin{figure*}
\includegraphics[angle=-90, scale=0.7]{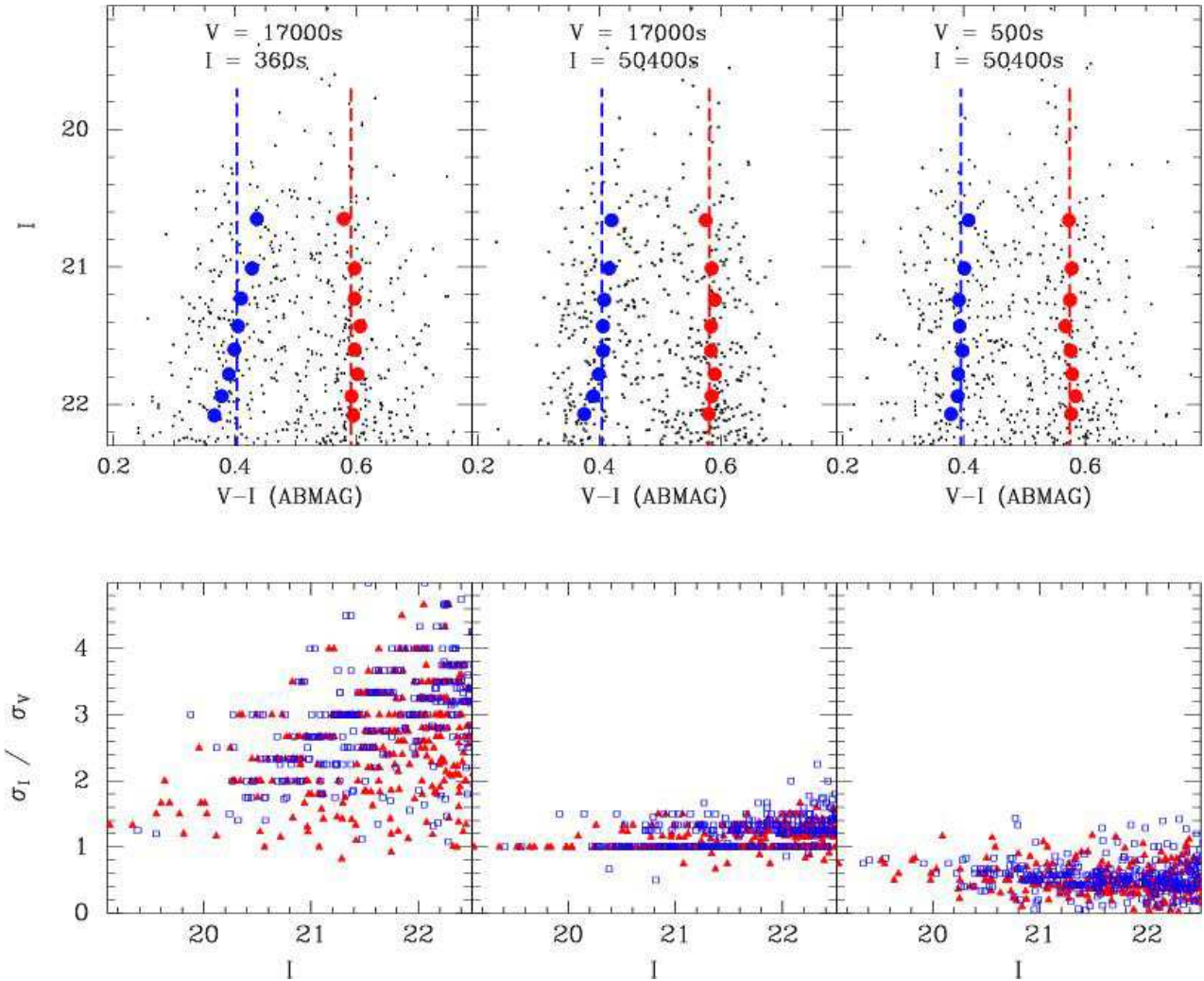}
\caption{ Top: V-I vs I CMDs of M87 GCs using data of different depths in the V and I band. The `blue-tilt' effect is significantly amplified when a deep V band observation is coupled with a shallow I band image, and almost negated when a shallow V data set is combined with a deep I image. Bottom: The formal Poisson uncertainties of GCs in the V and I band. The open squares indicate blue GCs and the filled triangles plot red clusters. Clearly larger relative uncertainties in the I band increase the `blue-tilt' effect in V-I vs. I CMDs.   }
\end{figure*}

	Globular cluster candidates were identified in the deep V and I images using the technique of Kundu et al. (1999). We measured the magnitudes and colors of the cluster candidates in both the deep images and a random set of single V and I band exposures using aperture photometry.  A 3 pixel aperture, 5-8 pixel sky, and aperture corrections and ABMAG zero points from Sirianni et al. (2005) were used for the photometry. The middle panels of Fig 7 plot the V-I vs. I CMD of the clusters in the deep images and the ratio of the Poisson uncertainties in the I and V band as a function of cluster magnitude. A small `blue-tilt' can be seen in the metal-poor clusters. The smaller color baseline (and hence the smaller relative difference in the size resolution) and the deeper data set conspires to produce a weaker `blue-tilt'. However we note that the surface brightness noise is an inherent astronomical signal (or noise), and irrespective of how deep a data set is it causes a small amplification in the `blue-tilt'. The left panels show that when a deep V image is combined with a single 360s I image the `blue-tilt' increases due to the larger noise in the I band photometry. On the other hand when a shallow 500s V band image is combined with a deep I band image the `blue-tilt' is weakened to a negligibly small effect. The magnitude of this effect changes depending on which particular short exposure image is chosen for this test, but the overall trend is consistent i.e. using a short I exposure amplifies the tilt and a short V exposure weakens it. This shows that variations in the photometric noise whether it be instrumental or astronomical in origin can significantly modify the properties of an apparent `blue-tilt'.

This effect of photometric noise can explain the apparent strengthening of the `blue-tilt' in the inner regions of galaxies reported in some studies (Mieske et al. 2006). The brighter galaxy background and the consequently higher absolute Poisson and SBF noise amplify systematic effects. Other physical effects may also play a subtle role. For example, effective sizes of globular clusters are known to correlate with galactocentric distance (e.g. van den Bergh 1994; Spitler et al. 2006). Although the fraction of distant GCs that are projected on to the inner regions of galaxies increases with decreasing galactocentric distance, large, low mass GCs projected on to the inner regions may not be detected because of the galaxy background. Moreover, high mass GCs in the inner regions of galaxies are preferentially destroyed due to dynamical friction. The increased fraction of high mass and large GCs at small projected galactocentric distances, and the relative smaller fraction of low mass and large GCs seen in the inner regions can enhance the observed `blue-tilt'.

Mieske et al. (2006) also suggest that the `blue-tilt' may be stronger in more massive galaxies. This can be explained by the more metal-rich and hence redder colors of the underlying stellar population in higher mass galaxies. This increases the relative photometric and surface photometry noise in the red filter in such galaxies. Conversely a small `red-tilt' is seen in the metal-rich clusters in g-z vs. g CMDs of the least massive galaxies because the underlying light is bluer and hence contributes relatively more noise to the photometry in the bluer filter. Moreover, both the `blue-tilt' and the mass-size correlation that is the underlying cause are primarily seen in the brightest, highest mass GCs in M87. Lower mass galaxies that are significantly less rich in globular clusters clearly have far fewer (if any) GCs populating such a high mass tail of the distribution.

\begin{figure}
\includegraphics[angle=-90, scale=0.4]{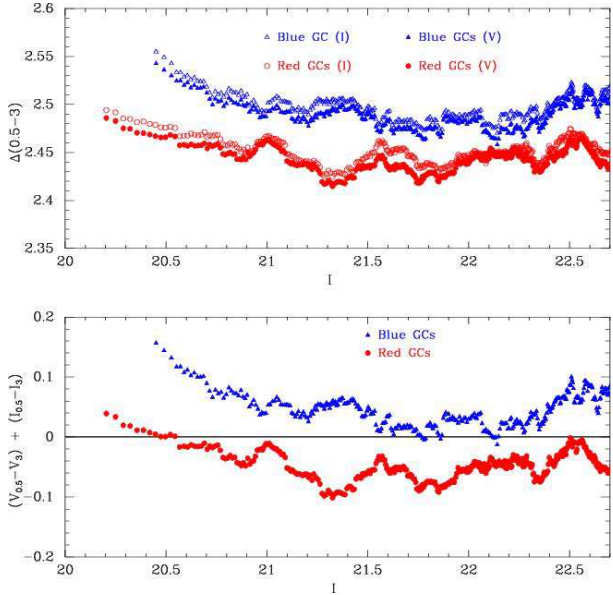}
\caption{ Top: The $\Delta$(0.5-3) size distributions of the GCs in the deep M87 V and I band data set plotted in Fig 7. Bottom: The (V$_{0.5}$-V$_3$) + (I$_{0.5}$-I$_3$) magnitudes of the red and blue GCs. The mass-size relationship for luminous GCs in M87 is more obvious than in Fig 4 due to the deeper data set.}
\end{figure}

	Following the method in $\S2.2$ we also measured the magnitudes of the cluster candidates in different apertures. As in Fig 4, Fig 8 shows the size distributions of the metal-rich and metal-poor clusters as a function of magnitude. The mass-size relationship at the high mass end of  both the metal-poor and metal-rich GCs is apparent. There is less scatter in the luminosity vs. size distribution for the fainter sources than in Fig 4 due to the better sampling and significantly higher S/N of this data set. The size distribution of the lower mass GCs {\it may} indicate a secular trend consistent with an anti-correlation with mass, as expected from cluster evaporation (Fall \& Rees 1977): But we hasten to add that any such claim requires a far more detailed analysis that is beyond the theme and scope of this paper. We note that the apparent correlated rise in GC sizes in both the blue and the red population near I$\approx$22.5 mag is purely coincidental and due to a handful of GCs with apparently large sizes in both sets. Nearly all the apparently large red GCs at that luminosity are in the vicinity of another bright object and hence likely have unreliable size estimates. The handful of large blue GCs at those luminosities however may indeed be large. But, this is not surprising if the larger mean sizes of the blue GCs as compared to red GCs is due to the projection of objects that are at larger galactocentric distances (Spitler et al. 2006). The scatter in the size distribution and the mass-size trend of blue GCs may be increased due to the consequently larger stochastic effects of sampling GCs that encompass a larger size distribution (due to the galactocentric distance vs. size correlation). Regardless of the nature of the luminosity vs. size relationship of low mass GCs the primary conclusion from our study is that there is indeed a small, but consistent correlation between the mass and size of luminous globular clusters. This is the primary cause of the apparent `blue-tilt' in high mass GCs. The direction of the uncertainty ellipses in the color-magnitude diagrams typically used to study the apparent `blue-tilt' serves to exacerbate this observational effect.

\subsection {Error is a Hardy Plant: `Blue-Tilts' in Other Galaxies}

We have shown that the features, and the magnitude, of the `blue-tilt' can be explained by a combination of the properties of the clusters, the host galaxy, and measurement uncertainties. Can this account for other published claims of `blue-tilts' in various galaxies? In this section we investigate apparent `blue-tilts' in some other galaxies.

	Both the Strader et al. (2006) and Mieske et al. (2006) studies have analyzed the GO-9401 data of Virgo cluster galaxies. However, both restrict the analysis of blue tilts in individual galaxies to the three most luminous (massive) candidates M49 (NGC 4472), M87 (NGC 4486) and M60 (NGC 4649). Consequently these are also the most GC rich galaxies in the sample, although NGC 4649 and NGC 4472 have fewer than half the number of GCs as M87 within the ACS field of view. According to our explanation above the apparent `blue-tilt' is primarily driven by the mass-size relationship of the most luminous GCs, so it is not surprising that `blue-tilts' can only be studied individually in the richest cluster systems that have a significant tail of luminous and large GCs that are at least partially resolved in HST images.

	Strader et al. (2006) report a $d(g-z)/dz$ slope of -0.043$\pm$0.01 for the blue GCs in M87 while Mieske et al. (2006) measure a slope of -0.042$\pm$0.015. It is important to stress that the uncertainties estimate the reliability of the respective algorithms used to measure the peak colors {\it and not} the uncertainties in the underlying data set. For the sparser NGC 4649 cluster system, using different magnitude cutoffs and sampling frequencies,  Strader et al. (2006) and Meiske et al. (2006) measure somewhat different but weaker slopes of -0.037 and -0.028$\pm$0.009 respectively. On the other hand both claim  that NGC 4472 does not show a `blue-tilt', with Mieske et al. (2006) measuring a slope of -0.008$\pm$0.024. This has been suggested as evidence of differences in the formation mechanisms of these galaxies, and has been argued as proof that the `blue-tilt' is real. However, in a recent study of a different NGC 4472 data set Lee et al. (2008) claim to find evidence of a `blue-tilt' in this galaxy. In order to explore this apparent discrepancy we re-analyzed the GO-9401 images of NGC 4472 studied by Strader et al. (2006) and Mieske et al. (2006) in a manner identical to the M87 study described above. Restricting the `blue-tilt' analysis to the same z=22 mag of the M87 study we indeed find a `blue-tilt' in this galaxy with a  $d(g-z)/dz$ slope of -0.025, similar to that of NGC 4649.

	The crucial difference between our estimate and the Mieske et al. (2006) study is the cutoff magnitude of the sample considered (Strader et al. 2006 do not present relevant numbers for NGC 4472). When we extend our analysis to the z$\approx$23.4 mag limit adopted by Mieske et al. (2006) we find a similarly negligible tilt. This is simply a reflection of the fact that the mass-size relation is only apparent for the brightest GCs in the GO-9401 data set (e.g. Fig 4). As discussed above the significantly larger scatter in the size-magnitude relation of the GO-9401 M87 data set plotted in Fig 4 as compared to the much deeper data shown in Fig 8 indicates the combined effects of lower signal-to-noise and poorer sampling of the PSF in the spatially undersampled data set of the GO-9401 images of Virgo cluster galaxies. This in turn adds noise to the photometry that is correlated to the size estimates. Due to the large GC population of M87, and hence the longer tail of luminous and larger GCs all studies consistently find a significant `blue-tilt'. But we note that even for M87 both Mieske et al. (2006) and Strader et al. (2006) find that estimates of the blue peak at some magnitudes are not consistent with the overall trend.  The quoted uncertainties in published values of the `blue-tilt' that quantify the reliability of the algorithms used to measure the peak colors underestimate the true uncertainties. Thus variations in such quoted values of the slope of the `blue-tilt' between NGC 4472 and the similarly rich NGC 4649 cluster population, based on an arbitrary faint cutoff magnitude do not provide any convincing evidence of differences in the `blue-tilt'. It simply reflects the fact that as compared to M87 a smaller fraction of the blue GCs in the sparser NGC 4472 and NGC 4649 cluster systems are well resolved.

More interestingly GCs in Virgo and Fornax galaxies are expected to be unresolved in ground-based images. As the mass-size relation discussed in $\S$2.2 is not expected be a factor in most ground-based studies the color-magnitude trend should be unaffected. Moreover, since ground-based GC studies typically mask out the bright inner regions of galaxies the SBF and noise effects should be reduced, and little or no `blue-tilt' is expected in such studies. The deep ground-based study of NGC 1399, the central galaxy in the Fornax cluster, by Dirsch et al. (2003) shows just such remarkably constant peaks in the colors of the metal-poor and metal-rich clusters in the metallicity sensitive C-T1 colors over many decades of luminosities. There is clearly no `blue-tilt' for GCs in the luminosity range -10.5$\lesssim$M$_{T1}$$\lesssim$-8.5, or -10$\lesssim$M$_{T1}$$\lesssim$8 in the Dirsch et al. (2003) data. Other ground-based studies of NGC 1399 by Forte, Faifer, \& Geisler (2007) and Ostrov, Forte, \& Geisler (1998) suggest a similar lack of a mass-metallicity relationship over a large luminosity range.

 Harris et al. (2006) argue in their HST-ACS based study of brightest cluster galaxies that there is a color-magnitude trend in the GCs brighter than M$_I$$\lesssim$-10.5 (M$_V$$\lesssim$-9.5) of their sample. It can be argued that since the Harris et al. (2006) study probes distant galaxies and probably does not resolve cluster sizes this is indicative of a true mass-metallicity effect. We note however that Harris et al. (2006) do claim that 20-30 of the brightest objects around  the closest galaxy of their sample NGC 1407 (23 Mpc) are partially resolved. Harris et al. (2006) argue that the sizes of the GCs in the cluster rich distant galaxies in their sample, which show the strongest evidence of the mass-metallicity effect, are not resolved. But if these luminous intermediate color clusters are larger than typical candidates, as is expected from the mass-size relation shown in Figs 4 \& 8, and is indeed observed by Strader et al. (2006) in their study of Virgo ellipticals, it is not at all clear that the brightest clusters are necessarily unresolved.

\begin{figure*}
\includegraphics[angle=0, scale=0.7]{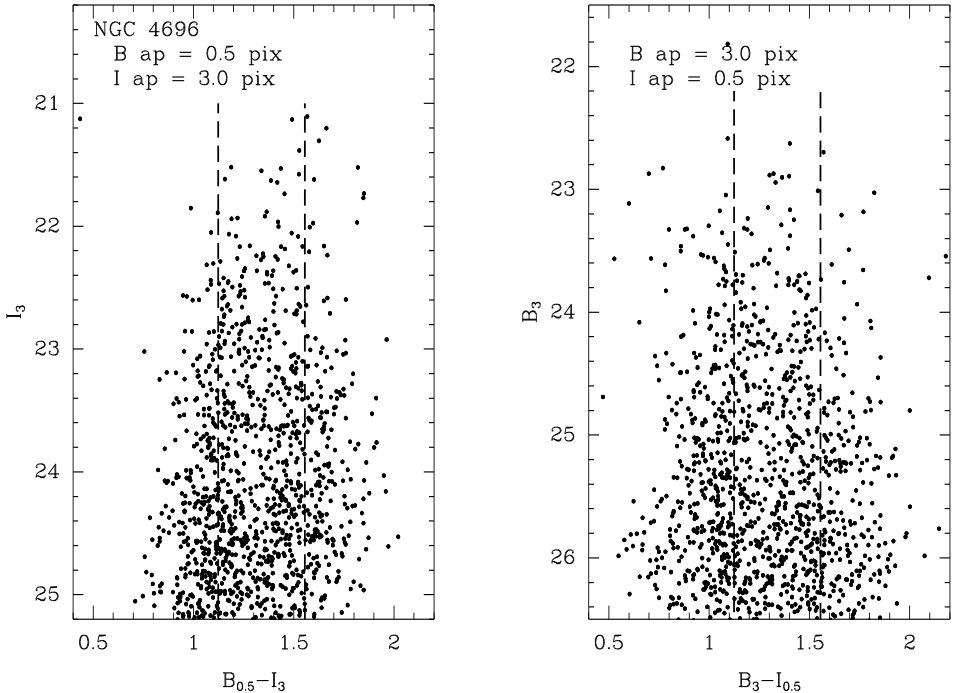}
\caption{ Left: The B$_{0.5}$-I$_3$ vs I$_3$ CMD of globular clusters in NGC 4696. The apparent trend towards redder colors for the brightest GCs is similar to that observed by Harris et al. (2006) in this distant, luminous, galaxy. Right: The B$_3$-I$_{0.5}$ vs. B$_3$ CMD of GCs in NGC 4696. The brightest GCs appear to trend bluer in this plot. The color distribution of the brightest GCs differs in the two representations of the CMD. This indicates that the apparent trend in the Harris et al. (2006) data is likely caused by a small GC mass-radius relationship and that the high mass GCs are at least partially resolved in HST images  of this 42 Mpc distant galaxy.  }
\end{figure*}

At a distance of 42 Mpc NGC 4696 is among the most distant galaxies in the Harris et al. (2006) study. It is also the galaxy with the richest globular cluster system and shows the most obvious signature of an apparent mass-metallicity trend of all the bright ellipticals in their sample according to Harris et al. (2006). We have re-analyzed the globular clusters in the 5440s F435W (B band) and 2320s (I band) GO-9427 images of NGC 4696 studied by Harris et al (2006). The cluster candidates were identified and colors and magnitudes measured in the manner described in $\S2.1$. We performed photometry with both a 0.5 pixel, and a 3 pixel radius aperture in order to investigate the possible effects of GC size. The left panel of Fig 9 shows the B$_{0.5}$-I$_3$ vs I$_3$ CMD for 0.5 pixel aperture B band, and 3 pixel I band photometry. The lack of objects in the top left part of the diagram and the apparent tendency of clusters brighter than I$\approx$23 to trend redder at higher luminosity is consistent with the Harris et al. (2006) analysis. This has been interpreted as evidence of metallicity enrichment by Harris et al. (2006). We note that the colors and absolute magnitudes shown here appear different from the Harris et al. (2006) values because all the photometry in this paper is in the ABMAG system, and not VEGAMAG. Furthermore, contamination of the sample is irrelevant for the bright GCs that is the focus of our attention here, as is also noted by Harris et al. (2006) themselves.

	The right panel of Fig 9 shows the  B$_3$-I$_{0.5}$ vs. B$_3$ CMD for 3 pixel B band and 0.5 pixel I band photometry. The structure of the CMD for the brightest clusters is significantly different from the left panel. Some bright GCs appear to have extremely blue colors between 0.5$<$(B-I)$<$0.8 mags. The bright GCs also appear to generally trend towards bluer colors at higher luminosities. We note that both panels in Fig 9 plot the same set of candidate objects fainter than I$_3$=21 with Poisson uncertainties less than 0.2 mags. {\it No color selection has been applied.} This is important because it eliminates the possibility that the relative difference in the photometric uncertainties in various aperture and filter combinations induces a selection bias in the color distribution of the brightest clusters. In fact the fainter GCs at I$_3$$\sim$24 (and corresponding B$_3$$\sim$25) appear to show less of a difference in the color mag structure than the brightest objects. These features of the CMDs strongly suggest that the sizes of the brightest GCs are partially resolved and the trends in the color-magnitude plane are observational effects.

	Our analysis implies that the mass-size trend is continuous all the way out to the most luminous GCs in a sample. Hence, NGC 4696, the galaxy with the richest globular cluster system in the Harris et al. (2006) study has a significant tail of high mass clusters with relatively large sizes. These are partially resolved even at 42 Mpc. As the apparent `blue-tilt' in one of the most distant galaxies in the Harris et al. (2006) sample and that in galaxies that are much nearer (e.g. Spitler et al. 2006, Strader et al. 2006) can all be explained by a mass-size relation it stands to reason that this explanation also applies to the other, primarily nearer, galaxies in the Harris et al. (2006) sample. The strength of an apparent `blue-tilt' in a particular galaxy depends on the distance of a particular galaxy and the richness of its cluster system, with nearby, populous globular cluster systems revealing the strongest effect.
But this does raise the interesting question of just how large some of these clusters are. This study provides some hints. Harris et al. (2006) point out that $\omega$ Cen like clusters with half-light diameters of 13 pc would be unresolved with the effective 25 pc FWHM at NGC 4696 distances. Thus some of the most luminous GCs apparently have half-light radii of tens of parsecs. Such luminous, large objects that populate the bright luminosity tail of the most cluster rich systems might not have counterparts in the sparse Milky Way system. Moreover, as we discuss in greater detail in $\S$2.7 the well known galactocentric distance - GC size relation, coupled with the larger spatial extent of these rich cluster systems predicts a population of such large objects.  We note that in good seeing conditions even ground-based images at Virgo/Fornax distances can have FWHMs that correspond to $\sim$40pc in the sky. The apparent intermediate colors of the very brightest GCs in ground-based studies of NGC 1399 (e.g. Dirsch et al. 2003) {\it may} indicate a slightly resolved metal-poor cluster population. Furthermore for galaxies with rich globular cluster systems that are closer than Virgo/Fornax a larger fraction of the most luminous GCs may be partially resolved even in ground-based images, leading to a small tilt when standard reduction techniques are applied. Such a distance dependent effect is clearly visible in ACS data: Spitler et al. (2006) find a stronger tilt in {\it both} the blue and the red GCs in NGC 4594 as compared to the Meiske et al. (2006) and Strader et al. (2006) study of the more distant Virgo cluster sample.

Some recent ground-based studies of galaxies at Virgo distances and beyond have claimed evidence of `blue-tilts'. However, many of the claims are actually inconsistent with the trends seen in HST data. For example the Lee et al. (2008) study of NGC 4472 not only suggests a `blue-tilt' for the blue GCs in NGC 4472 but a similar `blue-tilt' for the $\it red$ GCs in this galaxy, which is not observed in HST studies of Virgo galaxies. Moreover Lee et al. (2008) find a `blue-tilt' in the blue GCs of NGC 4649, which is apparently consistent with the HST studies discussed above. But they also find a `red-tilt' for the red clusters which is completely inconsistent with the same
analyses. In a recent study ground-based study of the distant Antlia galaxies NGC 3258 and NGC 3268 Bassino, Richtler, \& Dirsch (2008) claim that the bimodal peaks of the GC color distributions of both galaxies become unimodal at high luminosities and hence corroborates the conclusions of Harris et al. (2006). However, the Bassino et al. (2008) analysis actually finds that the color distribution for both galaxies becomes progressively {\it bluer} with increasing GC luminosity, which is exactly the {\it opposite} of the trend claimed by Harris et al. (2006).
 This is a reminder of the tenuous nature of the claims of such effects and of the difficulty in ascertaining the reality of a small shift in color over a large range in luminosity. Small systematic variations in observational parameters such as a varying PSF, problems in registering images or color correction terms, possible saturation of a few bright GCs etc. can easily conceal or induce an apparent tilt in ground-based images. As discussed above many apparent claims of mass-metallicity trends in ground-based studies are actually contradictory, but the aspects which support the accepted wisdom appear to be amplified.

We note here that even though the apparent trend of cluster mean color with the GC luminosity (mass) is extremely small on an absolute scale (a few hundredths of mag shift in color per magnitude in luminosity) part of the reason it has been taken at face value is that it is seen in the most luminous GCs. Such objects are presumed to have the most secure photometry. However, we have shown above that this is a mistaken assumption and the color distribution of the brightest GCs derived using standard photometric techniques is actually {\it less reliable} due to the systematic effect of GC size. It has been suggested (Harris 2006) that the similar nature of the structures derived from PSF and aperture photometry implies that the photometry is reliable. There are two problems with this line of argument. Firstly, PSF photometry {\it assumes} that the source is unresolved and well fit by the chosen PSF. For a slightly resolved source PSF photometry will systematically underestimate the luminosity of the source leading to exactly the same bias as aperture photometry. Moreover, PSF fitting is also performed within a fixed aperture. Although Harris et al. (2006) do not specify the aperture within which they fit the PSF they mention that the FWHM is 2.1-2.5 pixels and that they apply a uniform mean aperture correction between the PSF fitting radius and a 10 pixel radius aperture. We note that the `blue-tilt' is caused by the failure to account for size dependent variations in the fraction of the total GC light that falls {\it outside} the aperture used for typical photometry whether PSF or aperture photometry is performed (the total light inside the aperture is measured by either technique irrespective of the differences in the spatial distribution). In other words the PSF fitting technique described by Harris et al. (2006) is virtually identical to aperture photometry; Hence it is not at all surprising that it yields similar results, with similar biases, as aperture photometry.

This raises the interesting question of whether the sizes can be accounted for to yield a corrected CMD. We argue above that the undersampling and signal-to-noise limitations of the GO-9401, g and z band data set of M87 (and other galaxies in the GO-9401 project) limits the reliability of the GC size measurements, and hence the associated photometric corrections, for the majority of GCs. However, Meiske et al. (2006) claim to have simultaneously fit the size and magnitude of each GC based on the photometric technique briefly mentioned in the Appendix of Jordan et al. (2005). The fact that the `blue-tilt' is still observed in such apparently corrected data suggests that this technique has in fact not been successful in determining each quantity independently. One possible reason is that Jordan et al. (2005) fit the profile of the GCs within a 4 pixel radius aperture for the majority of their objects. As we argue above this will give results with the same biases as aperture photometry. More problematically, Jordan et al. (2005) simultaneous solve for the concentration, half-light and luminosity of each GC. While any number of parameters can be fit to any light profile to minimize the residuals the true test of any photometric technique is how well the input parameters can be recovered. In one test of the apparent core radii of M87 GCs measured by the GO-9401 team, Smits et al. (2006) showed that it is consistent with there being no signal in this particular data set.
 In fact Carlson \& Holtzman (2001) argue that even for HST data the concentration parameter and cluster size can be measured reliably and simultaneously only for very high signal-to-noise candidates, a standard that is not met by the vast majority of GCs in the GO-9401 Virgo data sets or the corresponding GO-10217 Fornax data.  Jordan et al. (2005) indicate that {\it on average} the half-light radius can be reliably measured in  their analysis of GO-9401 data. But it is well known that the half-light is much more stable than other correlated GC parameters such as the concentration index, that Jordan et al. (2005) fit simultaneously (Kundu \& Whitmore 1998; Carlson \& Holtzman 2001). We also note that the Jordan et al. (2005) tests indicate biases in even the measured half-light radii for objects that are either smaller or larger than typical GCs.
Since neither the GC lists nor the non-standard algorithm used by Jordan et al. (2005) is public the effects of the three parameter fits on other correlated GC parameters such as the luminosity and core radius and it's implications on other claimed trends is not known.

Another important issue is the behavior of the ACS PSFs in the two particular filters F475W (g) and F850LP (z) chosen for the GO-9401 (and the corresponding corresponding GO-10217 Fornax) survey. The F850LP filter is known to produce a large halo due to long wavelength scattering (e.g. Sirianni et al. 2005). Although this effect is difficult to model this likely has the same effect as broadening the F850LP PSF and hence increasing the contrast in the size resolution of the g and z band. Thus the `blue-tilt' is expected to be further amplified in g and z band studies. Moreover this long wavelength halo is known to be correlated to the color of the observed object. At the blue wavelengths sampled by the F475W filter there is similar (albeit smaller) wavelength dependent large angle scattering due to charge diffusion effects. Fig 9 of Sirianni et al. (2005) reveals that due to the shape of the aperture correction vs. effective wavelength curves these effects are additive and the relative sizes of the F475W and F850LP PSFs change with color. Clearly this will affect the relative light profiles of clusters of different colors and hence the `blue-tilt'.  The exquisitely well characterized color dependent PSFs required for correcting barely resolved sources in g and z are not available. However if the PSFs corresponding to a single color is adopted,  color dependent PSF effects may manifest itself in studies of GC sizes: There should be small GC color dependent  offsets between the sizes measured in the g and z filters (assuming that GCs have the same actual size in g and z). We note here that while such wavelength dependent PSF effects can amplify or dampen `blue-tilts' they cannot be the primary cause of the `blue-tilt' because that would require the shape of the PSF to vary with {\it magnitude}.

	The deep, well sampled V \& I ACS data presented in Figs 7 and 8 largely sidesteps most of these issues and can indeed be used to measure various GC parameters much more reliably. The entire 50 orbit data set has been analyzed in detail by Waters (2008) as part of his PhD thesis. The photometry of individual GCs has been corrected for size effects in the normal course of that study. This completely independent analysis does not reveal a `blue-tilt' in the size corrected magnitudes (Waters et al. in preparation). This paper explores in detail the causes of the apparent `blue-tilt' in M87 and other galaxies, and the effects of cluster sizes on other GC trends.

Finally we note that one possible way to probe the reliability of any observed color-magnitude trend for luminous GCs in any data set (which may not be deep enough, or the PSF well characterized enough, to reliably correct for the effect of GC sizes) is to perform the tests shown in Figs 3 and 9. If the structure of the color distribution changes when CMDs using different combinations of aperture photometry measurements are plotted then the trend is likely caused by the data reduction technique and not a physical effect. On the other hand a true physical effect should be impervious to the choice or combination of photometric apertures. Within the uncertainties of the present observations the evidence indicates that the colors of nearly all the GCs in the metal-poor population are likely the same. The very brightest objects  such as the apparently intermediate color GCs in ground-based studies of NGC 1399 might also be ultra compact dwarfs (UCDs) (e.g. Drinkwater et al. 2003) or particularly large GCs. We discuss these possibilities later in the paper.

 \subsection {Missing the Trees For a Forest? The Host Galaxy Mass - Globular Cluster Metallicity Trend}

We have shown that that the apparent color luminosity trend in the metal-poor clusters is likely caused by observational effects and does not reflect an underlying metallicity trend. Can similar issues affect analyses of globular cluster metallicity vs. galaxy mass?  An increase in the mean color (metallicity) of red GCs with the mass of their host galaxies has been shown convincingly in many different studies (e.g. Kundu \& Whitmore~2001a; Larsen  et al.~2001; Peng  et al.~2006; Strader  et al.~2006). A corresponding trend in the metal-poor populations has a much smaller slope in color space and is hence much less convincing. However the recent ACS studies of Strader et al. (2006) and Peng et al. (2006) claim a statistically significant evidence for the increase in the mean color (metallicity) of the metal-poor GCs with host galaxy mass. Using a specific color-metallicity transformation Peng et al. (2006) also claim that the galaxy mass - GC metallicity trend has a similar slope for both metal-rich and metal-poor clusters. Such an observation has important consequences on the galaxy assembly process, because it suggests that both the metal-poor and metal-rich cluster subsystems know about the galaxy they formed in, and the cluster formation mechanisms for both the metal-poor and metal-rich GCs are similar. But how reliable are these observational trends, and are they affected by observational effects such as the ones that produce the artificial `blue-tilt'?

The grayscale kernel density plot of the peak colors of the globular clusters in  Peng et al. (2006) (their Fig 4) shows the relative significance of the galaxy mass-GC metallicity trend in the red and blue GCs. While the correlation in the metal-rich peaks is clear there is no obvious trend in the blue GCs. However, after rejecting galaxies with unimodal GC colors Peng et al. (2006) argue that the peak colors reveal a statistically significant trend for the metal-poor GCs. 
Since the uncertainty in the locations of the blue peaks is similar to the range of colors spanned by the metal-poor peaks of various galaxies it is difficult to select a sub-sample of galaxies that is representative of the trend. 
In order to investigate this effect further we selected 5 galaxies for a pilot study, M87 (VCC 1316), NGC 4472 (VCC 1226), NGC 4372 (VCC 759), NGC 4550 (VCC 1619) and NGC 4387 (VCC 828). These galaxies have been selected because they have at least 50 GCs with clear bimodality in the Peng et al. (2006) analysis. They also span a range of mean colors for the blue clusters (based on Peng et al. 2006) and follow the overall galaxy mass - blue globular cluster color trend in Peng et al. (2006). The data for these Virgo cluster galaxies from GO-9401 have the exact same observing setup and exposure as the M87 observations presented in Fig 1. We analyzed these data sets in a manner identical to the M87 analysis described above.

\begin{figure}
\includegraphics[angle=-90, scale=0.6]{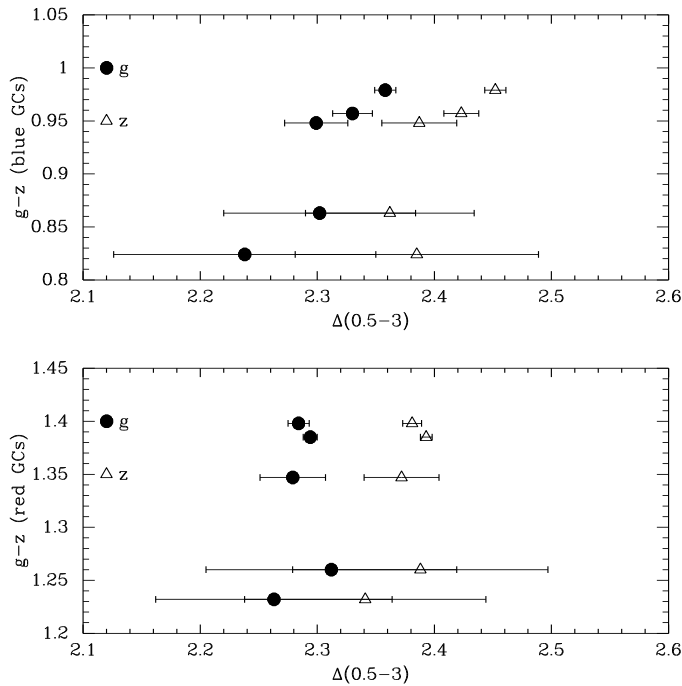}
\caption{ Top: The mean size of the blue population of GCs, as indicated by the $\Delta$(0.5-3) values, in five galaxies compared to the peak color of the blue GCs from Peng et al. (2006). Bottom: The corresponding plots for the red clusters. The top panel shows that the peak color of the blue GCs is correlated to the size of the clusters, while the bottom panel shows no such trend for the red clusters. The small galaxy mass - mean GC color trend for the metal-poor GCs is likely to be a consequence of this average size trend in blue GCs.  }
\end{figure}

Fig 10 plots the mean $\Delta$(0.5-3) size estimates in the g and z filters as a function of the g-z color of the red and blue peaks determined by Peng et al. (2006). Note that there is a one-to-one correlation between the mass of the host galaxy and the peaks colors of both the blue and red GCs determined by Peng et al. (2006) in this plot i.e. the galaxy mass increases along the y-axis. The top panel shows that the average size of the metal-poor GCs is larger in the more massive galaxies with apparently more metal-rich red {\it and} blue clusters. On the other hand there is no obvious trend in the mean sizes of the red GCs. We have shown in $\S$2.2 that adopting a constant aperture correction for GCs of different sizes as in Strader et al. (2006), Peng et al. (2006) and most other previous studies will bias the system with larger clusters to redder colors. This is likely the effect that has been interpreted as a galaxy mass - metal-poor GC metallicity trend in recent studies. We note here that we have plotted the average $\Delta{(0.5-3)}$ size estimates of only clusters brighter than an arbitrary limit of z=22 in order to minimize background contamination while ensuring a significant number of clusters in even the sparsest system. The broad correlations seen in Fig 10 are unaffected by the choice of the cutoff magnitude. Thus within the (typically underestimated) observational uncertainties this pilot study shows that there is no convincing evidence of a (galaxy) mass - GC metallicity relation for the metal-poor clusters, while there is a strong effect for the metal-rich clusters. 

	The trend of increasing metal-poor GC size with galaxy mass is Fig 10 may appear to be at odds with the Jordan et al. (2005) analysis which suggests that in fact the metal-poor GCs in low mass galaxies, with bluer galaxy colors, have larger sizes (their Fig 5). However, we note that Jordan et al. (2005) do not  plot the actual size of the GCs in their Fig 5 but the  `corrected' sizes adjusted for a size - galactocentric distance trend (which they apply in the form of a size vs. underlying galaxy z band surface brightness correction). The resulting correlation between galaxy mass and apparent cluster size derived by Jordan et al. (2005) may not have any bearing on the actual physical size of the GCs and the appropriate aperture corrections for these. We also note that although Jordan et al. (2005) apply a universal `correction' there is considerable galaxy to galaxy scatter in the GC size vs. z band galaxy brightness relationship. We also note that the average sizes of GCs reported in Table 1 of Jordan et al. (2005) appear to decrease with the mass of the host galaxy. While at first sight this might appear to inconsistent with the trends in Fig 10, we note that the sizes reported in Table 1 of Jordan et al. (2005) include both red and blue GCs. Since the fraction of the smaller red GCs increases with galaxy mass within the field of view of HST images (e.g. Peng et al. 2006) such a trend for the mean size of {\it all} GCs is not at odds with the correlations  in the {\it blue} GCs discussed above.

To first order the size of a globular cluster reflects the average density of 
the host galaxy internal to its radial distance. In as much as the $\Delta$0.5-3 pixel photometry reflects the effective size (or more accurately the tidal radius in this context) of the cluster Fig 10 implies that more massive galaxies are on average less dense. This is broadly consistent with the trend in effective galaxy densities seen in fundamental plane studies (e.g. Desroches et al. 2007).

 \subsection {What Error Leads Must Err: Other Implications of Globular Cluster Sizes}

In Figure 3 and the subsequent discussion we showed that the peak colors of both the red and the blue globular cluster subsystems can shift by few hundredths of a magnitude in different directions entirely because of a small mass-size relationship in GCs. We note the shifts in the colors arise from the minor differences in the PSFs in the two filters under consideration. The size dependent offsets between the true and measured magnitudes in each filter actually increase in the same direction and are therefore much larger than the apparent systematic color offsets. Thus, unless there is deep data for which an attempt can be made to correct the photometry for the sizes of individual clusters as in Waters et al (2006) and Waters (2008) any interpretation of small differences in the absolute magnitude distributions of GCs or the globular cluster luminosity function (GCLF) should be tempered by the consideration of this added uncertainty. 

The differences in the surface brightness distributions of various galaxies, the relative number of metal-rich to metal-poor GCs, the absolute sizes of the GCs, the total number of GCs (because rich cluster systems have a longer tail of resolved GCs), and the details of the photometric technique used add various degrees of additional uncertainty that must be accounted for when comparing galaxy to galaxy trends in GCLFs. The apparent host galaxy mass dependent trend in the peak of the GCLF and differences in the distribution of the high luminosity end of the GCLF that have been interpreted as evidence of systematic variations in the initial GC mass function by Jordan et al. (2006; 2007) should be considered in light of the added sources of uncertainty mentioned above. In fact these trends are not very significant within the likely underestimated error bars of Jordan et al. (2007). Moreover, another independent analysis of the {\it same data set} by Strader et al. (2006) found no significant galaxy mass dependent variation in the GCLF peak.

 We also note that the mean turnover magnitude of z$\approx$-8.4 mags in ABMAG for the galaxies in the Jordan et al. (2006; 2007) sample translates to a Vegamag of z$\approx$-8.95. However this is inconsistent with the V$\approx$-7.4 and I$\approx$-8.5 mags Vegamag turnover measured in many previous studies (e.g. Waters  et al.~2006 and references therein). The z and I bands are very close and stellar population models such as Bruzual \& Charlot (2003) and Maraston et al. (2005) predict I-z colors that at best account for half the difference between the I and z turnovers for old GCs. Despite such issues what is perhaps more amazing, as has long been noted, is just how remarkably constant the GCLF peak appears to be, and how challenging it has been to explain this theoretically.

 \subsection {A Molehill! A Mountain! A Hillock? Cluster Sizes and the Dwarf Globular Transition Objects  }

So far we have been exploring the effects of small changes in GC size, primarily with luminosity, on various GC properties. There is in fact another much stronger trend in cluster sizes: The half light radius of a GC in the MW is known to be roughly proportional to R$^{0.5}_{Galaxy}$ (e.g. Djorgovski \& Meylan 1994; van den Bergh 1994). Recent studies of nearby ellipticals reveal a similar trend  (e.g. Spitler et al. 2006). Thus the combination of a mass-size and a mass-galactocentric distance relation predicts a significant tail of large GCs in the outer halos of galaxies, especially in globular cluster rich ellipticals. This can have considerable impact both on the measurement of the integrated properties of these large objects and in the interpretation of the transition objects between globular clusters and dwarf galaxies.

In the past decade a class of stellar systems, dubbed ultra compact dwarfs, that appear too large to be GCs, and too compact to be normal dwarf galaxies has been identified  (Hilker et al. 1999; Drinkwater et al. 2003). However, distinguishing between a compact dwarf and an unusually massive GC is a tricky proposition. For example Drinkwater et al. (2003) identified 5 UCDs in the Fornax cluster on the basis of luminosity - velocity dispersion distributions. However, Hasegan et al. (2005) argue that four of the 5  Drinkwater et al. (2003) UCDs are actually consistent with the distribution of GCs on such a plane. Hasegan et al. (2005) suggest using additional criteria to compare the location of such objects with respect to the apparently distinct fundamental planes defined by globular clusters and dwarf galaxies. We point out below that  it is not at all clear that a single plane necessarily defines clusters everywhere.

The structure of a GC can be well well described by reduced isothermal King (1966) models which are defined by 3 independent parameters, the mass, a fiducial radius, and a concentration parameter. Observationally, the mass to light radius adds another parameter. Various studies have shown that the mass to light ratios of Milky Way GCs are roughly constant (e.g. McLaughlin 2000) providing one constraint. However recent studies by Noyola \& Gebhardt (2006; 2007) that show that the inner light profiles of Milky Way and other local group galaxy globular clusters differ significantly  from  previous ground-based measurements raise uncomfortable questions about the reliability of this assumption. The central surface brightness, which is a crucial factor in core mass to light estimates, measured by Noyola \& Gebhardt (2006) are up to 2 mags brighter than values reported in the literature. Moreover the uncertainty in MW cluster profiles casts doubts on the reliability of the inner light distribution (and related quantities such as central velocity dispersion) measured in typically relatively shallow images of distant compact objects.

A further correlation between the properties of Galactic GCs constrains the clusters to a relatively narrow plane  (Djorgovski 1995; McLaughlin 2000). While the exact parametrization of the constraint varies from author to author the half light radius is either explicitly, or indirectly, a factor in all definitions of this fundamental plane of clusters. In fact the second constraint on the King model parameters that helps define the plane (besides the constant mass to light ratio) is essentially the statement, or assumption, that the mass-size relation for GCs is negligible, encoded in different sets of parameters by different authors (Djorgovski 1995; McLaughlin 2000)

But, as discussed above there is in fact a mild mass-size relation for GCs, and more importantly a stronger galactocentric distance-size relationship.  Given that reduced isothermal models by definition depend on the local potential of the host galaxy, some correlation of GC properties with R$_{Galaxy}$ is to be expected. Globular cluster destruction studies (e.g. Murali \& Weinberg 1997; Vesperini 2000) have long recognized that the effects of GC destruction mechanisms vary with galactocentric distance, thus inducing radial trends in cluster parameters. But some numerical simulations suggest that size of a GC is relatively constant over most of its lifetime (e.g. Aarseth \& Heggie 1998; McMillan \& Hut 1994). It is also possible that the spatial variation in GC sizes is largely imprinted at birth, as is implied in some formation models (e.g. Ashman \& Zepf 2001).

	Irrespective of whether the GC half-light radius - galactocentric distance correlation is primarily driven by formation or evolutionary physics it is likely that the fundamental plane is only an appropriate representation of GCs within some limited galactocentric range. In fact, it has been shown that the scatter in the fundamental plane is reduced when GC properties are scaled to a fixed R$_{Galactocentric}$ (e.g. McLaughlin et al. 2000; Barmby et al. 2007). However, as the physical properties of GCs are defined by their fundamental and derived parameters such as the half-light radius and binding energy it is not clear for example what a scaled quantity such as binding energy signifies physically. Furthermore, when the fundamental planes of GCs and galaxies are considered for identifying transition objects the spatially uncorrected values are adopted. This leads to the situation where nearly all of the luminous GCs in NGC 5128 analyzed spectroscopically by Martini \& Ho (2004) deviate from the GC fundamental plane in Hasegan et al. (2005) and appear to be ultracompact objects. If nearly every bright GC in NGC 5128 studied in detail is inconsistent with the GC fundamental plane it is likely that it is the definition of the fundamental plane that needs to be revised.

The underlying issue is that the GC fundamental plane is primarily defined by the relatively sparse Milky Way cluster system. Moreover, often only some well observed subset is considered. For example there are few high mass GCs at large Galactocentric distances in the Milky Way. The low mass GCs in the outer regions are usually dismissed as outliers, thus suppressing any radial size trend. The typical GC with a 3 pc half-light radius in the Milky Way is at 5.4 kpc according to the van den Bergh (1994) correlation. By contrast the NGC 5128 GCs that appear to scatter away from the Milky Way GC fundamental plane in the Hasegan et al. (2005) analysis lie between 6-24 kpc {\it projected} radii, with a mean projected distance of 11.9 kpc. Even accounting for the different mass profiles of the Galaxy and NGC 5128 the half-light radii of the latter sample are expectedly larger. Moreover, as the tidal radius is also a function of the local potential (or R$_{galactocentric}$) it is not surprising that various correlations between  the properties of the NGC 5128 GCs differ from the MW clusters.

Since the NGC 5128 GCs appear to follow correlations that are similar to those of galaxies rather than GCs (Hasegan et al. 2005) in many scaling relations, identifying UCDs by comparing them with Milky Way clusters is less than convincing. A more secure way to determine whether such objects are truly distinct from GCs would be to compare such candidates with the properties of clusters at large galactocentric distances. In this context we note that NGC 2419, the bright Milky Way Cluster identified as having a size similar to the M87 transition objects by Hasegan et al. (2005) is 90 kpc from the Galactic Center.  While Hasegan et al. (2005) identify likely UCDs within the HST field of view, it is not clear whether these may be more distant objects projected at smaller galactocentric radii. We note that dynamical friction plays an increasingly significant role for higher mass objects in the inner regions of galaxies, especially in less massive galaxies. Thus, at small projected galactocentric distances it is more likely that the most massive candidates being observed are actually located at larger radial distances.

Globular clusters and dwarf galaxies differ in some important aspects, such as the lack of a dark matter component in the former. The transition objects likely hold important clues that help pinpoint the genesis of these differences. For example, it may be that while dwarfs trace small scale primordial fluctuations, GCs, even the highest mass ones that have masses similar to dwarfs, are formed in disks (e.g. Kravtsov \& Gnedin 2005). In such a scenario the location of the objects may provide additional insight into the formation and evolution of the two classes of objects. In a recent study Thomas, Drinkwater, \& Evstigneeva (2008)  test whether UCDs may be the threshed nuclear remnants of dwarf galaxies and encounter difficulties in explaining the observational evidence within this scenario. Given that nearly most UCDs lie within $\sim$100 kpc of a giant galaxy (e.g. Thomas et al. 2008) and the expected variation of the fundamental plane with galactocentric distance the possibility that these objects are large and massive globular clusters, and not necessarily stripped dwarfs needs to be explored in greater depth.

\section{Conclusions}

In this paper we have investigated the underlying causes of a number of trends in GC properties that have recently been identified using HST imaging data. While many of these correlations are quite weak, they have been observed by a number of independent groups, and interpreted as evidence of cosmological effects. One of these is a small color-magnitude trend in the blue, metal-poor globular clusters in a number of galaxies, in the sense that more luminous clusters appear redder. This apparent mass-metallicity relationship, dubbed the `blue-tilt', has been construed as evidence of self enrichment in the largest clusters. 

We show in Figure 3 that both the magnitude, and even the direction, of the apparent trend can be changed by adopting different combinations of aperture radii for the photometry in each filter. Figures 4 \& 8 indicate the reason why. There is a small mass-size trend for globular clusters. Thus the fraction of light that is measured within any fiducial aperture varies with cluster luminosity. This, coupled with filter dependent differences in the point spread function induces a color-magnitude trend, particularly in the larger, metal-poor, blue globular clusters. Figure 9 shows that even for distant galaxies like NGC 4696 (at 42 Mpc) the mass-radius relationship coupled with the spatial resolution of the HST is enough to induce a color-magnitude correlation in the CMD.

We show in Figure 5 that the correlated color and magnitude uncertainties due to both Poisson noise and surface brightness fluctuations tend to amplify the `blue-tilt', particularly for the metal-poor clusters, when the redder of the two filters is used to construct the magnitude axis of a color-magnitude diagram. Conversely, using the bluer filter negates the tilt. Figures 6 \& 7 reveal the significant effects of noise on the magnitude of the tilt. This mix of often overlooked physical properties of clusters, and the effect of background noise can explain other reported correlations of the apparent mass-metallicity trend of metal-poor GCs with host galaxy mass and location in the galaxy. More massive galaxies are more metal-rich, and hence redder, thus systematically increasing the amplitude of the observational trend. The brighter inner regions of galaxies also obviously contribute larger photometric noise. Thus within the uncertainties of the observations there is no believable evidence for a mass-metallicity trend in metal-poor halo GCs, and no obvious signature of self enrichment in high mass clusters. 

Various recent HST studies have also suggested a small trend in the peak color of the metal-poor cluster sub-population with host galaxy mass. It has been argued that this trend, which is much weaker than the galaxy mass-globular clusters color correlation seen for the metal-rich clusters, implies the old, blue, metal-poor clusters retain knowledge of the larger galactic halo in which they formed. We show in Fig 10 that the average size of the metal-poor clusters in the low mass galaxies with apparently bluer (and hence more metal-poor) clusters on average are smaller than corresponding GCs in high mass galaxies. Such a correlation between the mean cluster size and host galaxy mass will induce precisely the small photometric trend between cluster color (metallicity) and galaxy mass when common photometric techniques are applied. Thus our studies show that within the uncertainties of the present data sets the old, blue, metal-poor GCs in the halos of galaxies have remarkably similar colors (metallicities) everywhere. We also point out that the overlooked trends in cluster sizes can affect the various other reported trends in cluster properties such as the apparent correlation of the peak of the globular cluster luminosity function with galaxy mass.

Finally we also point out that the well established trend of increasing globular cluster size with the distance from the center of its host galaxy has important implications on the commonly adopted fundamental plane of globular clusters and the search for ultra compact dwarfs. The GC fundamental plane is largely defined by clusters in the Milky Way that are within a rather limited range of Galactocentric distance. Given the galactocentric trend in cluster properties, combined with the mass-radius correlation for the high mass clusters, the high mass GC tail at large galactocentric distances are expected to have the largest effective sizes. Although such candidates are usually identified as UCDs it is not at all clear that objects are inconsistent with the {\it local} definition of the GC fundamental plane. We suggest that the spatial distribution of both globular clusters and UCD candidates need to be considered in order to identify possible candidates that might be truly distinct from the high mass tail of globular clusters.

I gratefully acknowledge support for this research from HST grants AR-11264, AR-09208 and GO-10529 and NASA-LTSA grant NAG5-12975. I thank Steve Zepf, Chris Waters, Enrico Vesperini, \& Kathy Rhode for many very helpful discussions.

\end{document}